\documentclass[prx,a4paper,twocolumn,nofootinbib,superscriptaddress]{revtex4-2}

\usepackage{amsfonts,amsmath,amssymb,bm}
\usepackage{graphicx}

\newcommand{\ket}[1]{\vert#1\rangle}
\newcommand{\bra}[1]{\langle#1\vert}
\newcommand{\pket}[1]{\vert#1\rangle}

\newcommand{\sx}{\sigma^x}
\newcommand{\sz}{\sigma^z}
\newcommand{\hsx}{\hat\sigma^x}
\newcommand{\hsz}{\hat\sigma^z}

\newcommand{\gid}{\mathrm{id}}

\begin{document}

\title{Entanglement order parameters and critical behavior 
for topological phase transitions and beyond}

\author{Mohsin Iqbal}
\affiliation{Max-Planck-Institute of Quantum Optics,
Hans-Kopfermann-Stra\ss{}e 1, 85748 Garching, Germany, and\\
Munich Center for Quantum Science and Technology,
Schellingstra\ss{}e~4, 80799 M\"unchen, Germany}
\author{Norbert Schuch}
\affiliation{Max-Planck-Institute of Quantum Optics,
Hans-Kopfermann-Stra\ss{}e 1, 85748 Garching, Germany, and\\
Munich Center for Quantum Science and Technology,
Schellingstra\ss{}e~4, 80799 M\"unchen, Germany}
\affiliation{University of Vienna, Faculty of Physics, Boltzmanngasse
5, 1090 Wien, Austria, and\\
University of Vienna, Faculty of Mathematics, Oskar-Morgenstern-Platz 1, 1090 Wien, Austria}

\begin{abstract}
Order parameters are key to our understanding of phases of matter. 
Not only do they allow to classify phases, but they also enable the study of phase
transitions through their critical exponents which identify the universal
long-range physics underlying the transition.  Topological phases are
exotic quantum phases which are lacking the characterization in terms of
order parameters. While probes have been developed to identify such
phases, those probes are only qualitative in that they take discrete values, and
thus provide no means to study the scaling behavior in the vicinity of phase
transitions. 
In this paper, we develop a unified framework based on variational tensor
networks (infinite Projected Entangled Pair States, or iPEPS) for the
quantitative study of both topological and conventional phase transitions
through \emph{entanglement order parameters}. To this end, we employ tensor
networks with suitable physical and/or entanglement symmetries encoded,
and allow for order parameters detecting the behavior of \emph{any}
of those symmetries, both physical and entanglement ones. On the one hand,
this gives rise to entanglement-based order parameters for topologically
ordered phases.  These topological order parameters allow to
quantitatively probe the behavior when going through topological phase transitions
and thus to identify universal signatures of such transitions. We apply
our framework to the study of the Toric Code model in different magnetic
fields, which along some special lines maps to the (2+1)D Ising model. Our
method identifies 3D Ising critical exponents for the entire transition,
consistent with those special cases and general belief.  However, we in
addition also find an unknown critical exponent $\beta^*\approx 0.021$ for one
of our topological order parameters. We take this -- together with known
dualities between Toric Code and Ising model -- as a motivation to also apply
our framework of entanglement order parameters to conventional phase
transitions. There, it enables us to construct a novel type of disorder
operator (or disorder parameter), which is non-zero in the disordered
phase and measures the response of the wavefunction to a symmetry twist in
the entanglement. We numerically evaluate this disorder operator for the
(2+1)D transverse field Ising model, where we again recover a critical
exponent hitherto unknown in the (2+1)D Ising model, $\beta^*\approx
0.024$, consistent with the findings for the Toric Code.
This shows that entanglement order parameters can provide additional
means of characterizing the universal data both at topological and
conventional phase transitions, and altogether demonstrates the
power of this framework to 
identify the universal data underlying the transition.
\end{abstract}

\maketitle

\section{Introduction}

Symmetries play a central role in modern physics. In particular, they are
the key to understand the way in which many-body systems, both classical
and quantum, organize themselves into different phases, a problem central
to condensed matter physics, high-energy physics, and beyond.  To this
end, one needs to consider the full set of symmetries of the interactions
which describe  a system at hand, and study whether its state obeys the same
symmetries or chooses to break some of them. This can be captured through
local order parameters which are chosen such as to detect a breaking of
the symmetry. The understanding in terms of symmetries and order
parameters, however, does not only enable us to classify the ways in which
many-body systems can order, but it moreover allows to quantitatively
assess how the system behaves as it undergoes a phase transition, which
forms the 
heart of Landau theory. Indeed, the scaling behavior of the order
parameter in the vicinity of a phase transition allows to extract the
\emph{universal} features of the transition, that is, the fingerprint
of its long-range physics; it is a most notable fact that phase
transitions in scenarios such different as liquid-gas or magnetic
transitions fall into the same few universality classes, which in turn
allows to use effective field theories to capture the universal long-range
physics.

Topological phases are zero-temperature phases of quantum many-body
systems which fall outside of the Landau
paradigm~\cite{wen:book,fradkin2013field}. They exhibit ordering,
witnessed e.g.\ by a non-trivial ground space degeneracy and excitations
with a non-trivial statistics (``anyons'').  Yet, those ground states, and
thus the topological phase itself, cannot be characterized by any
local order parameter. Instead, other probes for identifying topologically
non-trivial states have been developed, such as a universal constant
correction $\gamma$ to the area-law scaling of the entanglement entropy,
$S(A)=c|A|-\gamma$~\cite{kitaev:topological-entropy,levin:topological-entropy},
features of the entanglement spectrum~\cite{cirac:peps-boundaries}, or
properties extracted from a full set of ``minimally entangled'' ground
states which carry information about the statistics of the
excitations~\cite{moradi:s-matrix}.

Yet, all these probes for topological order suffer from a severe
shortcoming as compared to conventional order parameters: On the one hand,
conventional order parameters allow to \emph{identify} the phase at hand
-- they are \emph{qualitative order parameters}. But at the same time,
they also allow to \emph{quantitatively} study the behavior of the system
as it undergoes a phase transition, and to extract information about the
universal properties of the transition -- they are \emph{quantitative
order parameters}. While fingerprints for topological order such as the
topological correction $\gamma$ or anyon statistics are qualitative order
parameters for topological phases, they can only take a discrete set of
values by construction and thus cannot be used for a \emph{quantitative}
study of topological phase transitions.  This leaves the quantitative
study of topological phase transitions wide open, with information about
the underlying universal behavior limited to cases where
exact~\cite{trebst:tcode-phase-transition} or
approximate~\cite{tupitsyn:tcode-multicritical} duality mappings to other
known models can be devised, or where universal signatures can be extracted
from the scaling of the bulk gap~\cite{dusuel:tc-w-field} or the
CFT structure of the full entanglement spectrum of the 2D bulk at
criticality~\cite{schuler:ising-star}.

In this paper, we develop a framework for the quantitative study of
topological phase transitions through order parameters based on tensor
networks, specifically
iPEPS~\cite{verstraete:2D-dmrg,jordan:ipeps,bridgeman:interpretive-dance}. Given a
lattice model $H$, our method uses variationally optimized iPEPS
wavefunctions to construct order parameters
which characterize the topological features of the system, namely the
behavior of the topological quasi-particles (anyons) and the way in which
they cease to exist at the phase transition, that is, their condensation
and confinement. Unlike other signatures of topological order, these
order parameters vanish continuously as the phase transition is approached
and thus allow for the extraction of critical exponents which enable the 
microscopic study of topological phase transitions and the
verification and identification of their universal behavior.

We apply our framework to the study of the Toric Code model in a
simultaneous $x$ and $z$ magnetic field, where we use it to extract
different critical exponents which characterize the transition. On the one
hand, we recover the anticipated 3D Ising critical exponents $\beta$ (for
the order parameter) and $\nu$ (for diverging lengths), consistent with
previous evidence found for the 3D Ising universality
class~\cite{trebst:tcode-phase-transition,tupitsyn:tcode-multicritical,dusuel:tc-w-field}.
For the order parameter for deconfinement, however, we find a new and yet
unknown critical exponent $\beta^*\approx 0.021$.  Our framework thus
allows to extract the universal signatures of topological phase
transitions, but even goes further and provides access to additional
critical exponents.

The observation of a yet unknown critical exponent, together with the
well-known duality mapping between the Toric Code with a pure $x$ or $z$
field and the (2+1)D transverse field Ising model, motivates us to
investigate whether similar techniques can also be used to set up disorder
parameters for conventional phase transitions, such as for the (2+1)D
Ising model, and whether those exhibit those unknown critical exponents as
well. 

We therefore consider symmetry breaking phase transitions, which we simulate
variationally using iPEPS with the global symmetry encoded in
the tensor. We propose to use the response of the variational wavefunction
to the insertion of a ``symmetry twist'' on the entanglement degrees of
freedom as a disorder parameter, as we show that a non-zero value
implies being in the disordered phase. We study the proposed
disorder parameter numerically for the (2+1)D Ising model, and find a
critical exponent $\beta^*\approx 0.024$ (consistent with the Toric Code
result up to numerical precision), in agreement with the expected duality
mapping.  Our construction therefore constitutes a novel way to define
disorder parameters for conventional phases, which provide a new tool to
extract additional signatures of universal behavior at criticality from
the system. Notably, this construction intrinsically relies on the
description of the system in terms of symmetric PEPS, which gives access
to properties which cannot be captured in a direct way by probing the physical
degrees of freedom alone.

In order to achieve the goals of the paper, we build on a number of
ingredients. First, we exploit that iPEPS form a powerful framework for
the simulation of strongly correlated quantum spin systems, based on the
description of a complex entangled many-body wavefunction in terms of
local tensors which flesh out the interplay of locality and entanglement,
and we make use of the powerful variational algorithms developed for
iPEPS~\cite{li:ttn_bethe_simple_update,phien:ipeps-update-algorithms,corboz:iPEPS-CTM,vanderstraeten:iPEPS-gradient}.
Next, we exploit the key role played by entanglement symmetries in
describing topologically ordered systems: While these symmetries had
originally been identified in explicitly constructed model wavefunctions
with topological
order~\cite{schuch:peps-sym,schuch:rvb-kagome,buerschaper:twisted-injectivity,sahinoglu:mpo-injectivity},
they have recently also been found to show up in variationally optimized
wavefunctions for topologically ordered
systems~\cite{crone:peps-tcode-detectZ2}; they thus constitute the right
structure for the description of topologically ordered systems.  We thus
impose the corresponding symmetries when variationally optimizing the
iPEPS tensor.  Next, these symmetries are known to allow to model anyons
and study their behavior in explicitly constructed wavefunction
families~\cite{schuch:peps-sym,buerschaper:twisted-injectivity,bultinck:mpo-anyons,haegeman:shadows,iqbal:z4-phasetrans,iqbal:breathing-kagome,iqbal:rvb-perturb}.
A key step of our work is to show that it is possible to generalize this
description to the case of variationally optimized iPEPS. In particular,
this requires a careful consideration of the way in which order parameters
are constructed \emph{solely} based on the symmetries present, without
\emph{any} further information at hand. While this seems contrived for
regular order parameters (where the full Hamiltonian and its dependence on
external parameters such as magnetic fields is known) and for explicitly
constructed PEPS model wavefunctions (where the full tensor and its
parameter dependence are given explicitly), this turns out to be crucial
for variationally optimizied iPEPS, where we have no information available
but the symmetry itself; a significant part of the manuscript deals with
this discussion.

The remainder of the paper is structured as follows: In
Sec.~\ref{sec:topo-opar}, we develop our framework for the construction of
order parameters in topological phases. 
In Sec.~\ref{sec:tc-field}, we apply our method to the in-depth study of
the Toric Code model in different magnetic fields.  Finally, in
Sec.~\ref{sec:discussion}, we discuss some further aspects of the method, before
concluding in Sec.~\ref{sec:conclusion}.

\section{Construction of topological order parameters\label{sec:topo-opar}}

In this section, we describe how to construct and measure topological
order parameter using iPEPS. We start in Sec.~\ref{sec:2:peps} with an
introduction to iPEPS, a discussion of entanglement symmetries, and the
way in which those symmetries underly topological order and how they can be used
to construct anyonic operators at the entanglement level.  In
Sec.~\ref{sec:2:anyon-behavior} we discuss the different physical behavior
which those anyonic operators can display, and their relation to the
topological phase the system exhibits. 

The following two sections, \ref{sec:2:opar-qual} and
\ref{sec:anyopar-quantitative-and-gauge}, form the theoretical core of
the construction of topological order parameters: We develop the framework
of how to use anyonic operators to construct order parameters. The key
challenge is that this
construction must be based on the weakest possible assumption, namely that we only
know about the symmetry of the model at hand, without any other information
about the problem. This is since we describe the system by variationally
optimized iPEPS tensors on which we only impose the entanglement symmetry
-- thus, the way the symmetry acts is the only information which we can
be certain about, while all other degrees of freedom are subject to
arbitrary gauge choices. While such a situation seems contrived
in the case of an actual model where a concrete Hamiltonian is given, the
study of order parameters based solely on the 
underlying symmetry can nevertheless
be discussed in that general scenario, where it provides insights on their
own right. Specifically, in
Sec.~\ref{sec:2:opar-qual} we discuss how from symmetry considerations, we
can connect anyonic order parameters to conventional and string order
parameters in one dimension, and how symmetries underly the construction
of the latter; and in Sec.~\ref{sec:anyopar-quantitative-and-gauge}, we
discuss the additional obstacles which appear when transitioning to the
case where we want to use order parameters for the quantitative study of
phase transitions. There, knowledge of the symmetry alone seems
insufficient  due to the free (and a priori random) gauge degrees of
freedom, and we explain how this can be overcome by constructing
order parameters which are gauge invariant, as well as through the
introduction of suitable gauge fixing procedures.

Finally, in Sec.~\ref{sec:subsec:recipe}, we provide a succinct and
detailed technical recipe for how to measure topological order parameters in
practice.

\subsection{iPEPS, entanglement symmetries, and topological
order\label{sec:2:peps}}

We start by introducing infinite Projected Entangled Pair States
(iPEPS)~\cite{verstraete:2D-dmrg,jordan:ipeps,bridgeman:interpretive-dance}.
For simplicity, we restrict to square lattices; other geometries can
be accommodated either by adapting the lattice geometry or by
blocking sites. We denote the physical dimension per site (possibly
blocked) by $d$. An iPEPS of \emph{bond dimension} $D$ is given by a
five-index tensor 
\begin{equation}
A\equiv A^i_{\alpha\beta\gamma\delta}=
\quad\raisebox{-0.7cm}{\includegraphics{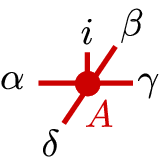}}\quad,
\end{equation}
with \emph{physical index} $i=1,\dots,d$, and \emph{virtual indices}
$\alpha,\beta,\gamma,\delta=1,\dots,D$.  It describes a wavefunction on an
infinite plane by arranging the tensor on a square grid and contracting
connected indices (that is, identifying and summing over them), 
depicted as
\begin{equation}
\label{eq:peps-contraction}
\raisebox{-.7cm}{\includegraphics{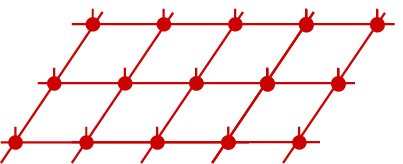}}\ .
\end{equation}
More formally, this contraction should be thought of as placing some
suitable boundary conditions at the virtual indices at the boundary and
taking those boundaries to infinity; numerically, this amounts to
convergence of bulk properties independent of the chosen boundary
conditions (except for possibly selecting a symmetry broken sector).

iPEPS form a powerful variational ansatz, as their entanglement structure
(built up through the contraction of the virtual indices) is well suited
to describe low-energy states of correlated quantum many-body systems, and
there exists a range of algorithms to find the variationally optimal state
for a given
Hamiltonian~\cite{li:ttn_bethe_simple_update,phien:ipeps-update-algorithms,corboz:iPEPS-CTM,vanderstraeten:iPEPS-gradient}.
At the same time, they can be used to exactly capture a range of
interesting wavefunctions, in particular renormalization fixed point
(RGFP) models with (non-chiral) topological order, as well as models with
finite correlation length through suitable deformations of the RGFP
models.

A key point of the PEPS ansatz is that there is a \emph{gauge ambiguity}:
Two tensors which are related by a gauge
\begin{equation}
\label{eq:peps-gauge-dof}
\raisebox{-0.7cm}{\includegraphics{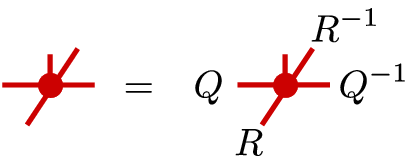}}\ ,
\end{equation}
(with \emph{gauges} $Q$ and $R$) describe the same wavefunction, as the
gauges cancel in the contraction~\eqref{eq:peps-contraction}. In
particular, for PEPS which have been obtained from a variational
optimization rather than having been constructed explicitly -- that is,
those which are
at the focus of this work -- we cannot assume a specific gauge, and
picking a suitable gauge will be of key importance later on.

PEPS models with topological order are characterized by an
\emph{entanglement symmetry} which is closely tied to their topological
features. This symmetry shows up in all known model wavefunctions with
topological order, but has recently also been found to appear in
variational optimized tensors, and is thus naturally linked to topological
order~\cite{schuch:peps-sym,schuch:rvb-kagome,buerschaper:twisted-injectivity,sahinoglu:mpo-injectivity,crone:peps-tcode-detectZ2}.
In the case of quantum doubles of finite groups
$G$~\cite{kitaev:toriccode} (which will
be the focus of this work), this entanglement symmetry is given
by
\begin{equation}
\label{eq:G-injective}
\raisebox{-0.7cm}{\includegraphics{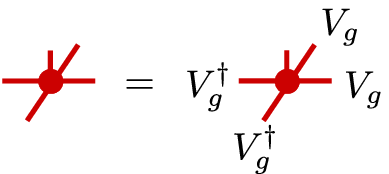}}\ ,
\end{equation}
where $V_g$, $g\in G$, is some unitary representation of
$G$~\cite{schuch:peps-sym}. (In the
graphical calculus, the $V_g$ are understood as $2$-index tensors which are
accordingly contracted with the virtual indices.)
Eq.~\eqref{eq:G-injective} implies a ``pulling through'' property: Strings
formed by $V_g$ (or $V_g^\dagger$, depending on the relative orientation
of the string and the lattice) can be freely deformed,%
\footnote{By correlating the actions on different links in the pulling through
condition in the form of a Matrix Product Operator, this framework can be
extended to encompass all string-net
models~\cite{sahinoglu:mpo-injectivity,bultinck:mpo-anyons}.}
 e.g.
\begin{equation}
\label{eq:pulling-through}
\raisebox{-0.6cm}{\includegraphics{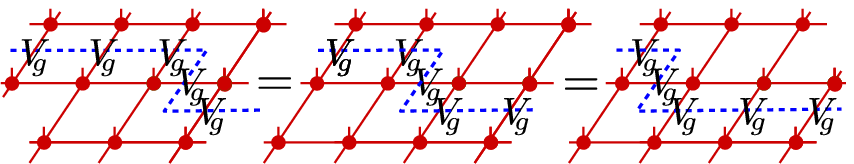}}
\end{equation}
For simplicity, in the following we will denote the $V_g$'s (or
$V_g^\dagger$) by blue dots, if needed labelled by placing the group
element $g$ next to it.

Restricting to tensors with a fixed symmetry \eqref{eq:G-injective}, as we
will do in our variational simulations, also induces a corresponding
symmetry constraint on the gauge degrees of freedom
\eqref{eq:peps-gauge-dof}: In order for the symmetry condition
\eqref{eq:G-injective} to be preserved, we must have that 
\begin{equation}
V_g Q V_g^\dagger = Q\mbox{\quad and\quad} V_gRV_g^\dagger = R\ .
\end{equation}

\begin{figure}
\includegraphics{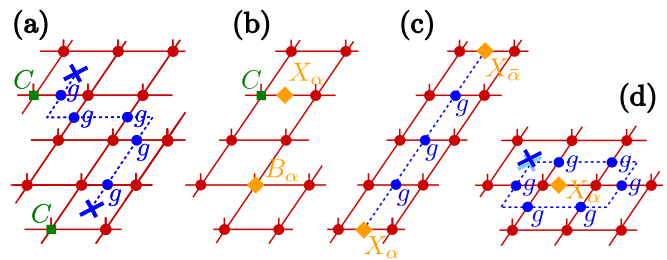}
\caption{\textbf{(a)} Strings of symmetry operations $g\equiv V_g$,
Eq.~\eqref{eq:G-injective}, possibly dressed by trivially transforming
endpoints $C$, form pairs of magnetic fluxes.  \textbf{(b)} Objects which
transform as non-trivial irreducible representations $\alpha$ under $V_g$ form
electric excitations, such as site tensors $B_\alpha$, or matrices $X_\alpha$
placed on links; again, they can be dressed with some trivially
transforming tensor $C$.  \textbf{(c)} A general pair of anyonic
excitations, as used in this work to study anyon condensation and
deconfinement.  \textbf{(d)} Braiding as described in the language of
entanglement symmetries \eqref{eq:G-injective} (see text).}
\label{fig:anyons}
\end{figure}

As it turns out, condition \eqref{eq:pulling-through} is closely tied to
topological order; in the following, we will focus on the case of Abelian
groups $G$ for simplicity.  First, we can use condition
\eqref{eq:pulling-through} to for instance parametrize a
ground space manifold with a topological degeneracy, by wrapping strings
of $V_g$ around the torus -- as those strings are movable, they cannot be
detected locally.\footnote{Strictly speaking, this is only rigorously true
for parent Hamiltonians which check the tensor network structure locally.}
Second, a string with two open ends -- see Fig.~\ref{fig:anyons}a -- allows to
describe paired excitations: While the string itself can be moved using
\eqref{eq:G-injective} and is
thus not detectable, its endpoints (which are plaquettes with an odd
number of adjacent $V_g$'s) cannot be moved, and we would thus expect them
to be detectable; these correspond to \emph{magnetic} excitations. On the other
hand, replacing a tensor by one with a non-trivial transformation property
\begin{equation}
\raisebox{-0.85cm}{\includegraphics{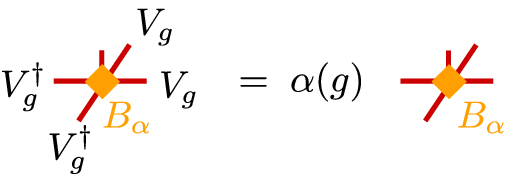}}
\end{equation}
where $\alpha(g)$ is an irreducible representation of $G$ (or,
alternatively, placing a matrix $X_\alpha$ with transformation property
\begin{equation}
\label{eq:X-irrep-phase}
V_g X_\alpha V_g^\dagger = \alpha(g) X_\alpha
\end{equation}
on a bond) -- see Fig.~\ref{fig:anyons}b -- also yields a topological excitation: As it carries a total
irrep charge under the action of $V_g$, it must come in charge-neutral
pairs on a torus (or otherwise be compensated by the boundary conditions).
Objects of this form are \emph{electric} excitations. For both 
these types of excitations, or combinations of electric and magnetic excitations
(``dyons''), we can additionally dress the endpoint with a trivially
transforming tensor $C$ (i.e., one which satisfies \eqref{eq:G-injective}), e.g.\ to
create an exact energy eigenstate. The most general pair of excitations
(without the dressing) is shown in Fig.~\ref{fig:anyons}c.

When seen on the entanglement degrees of freedom, these objects carry all properties
expected from anyonic excitations. They can only be created in pairs, and
if we assume for a moment that we have a way to move and probe them,
they exhibit precisely the statistics of the anyons in the double model $D(G)$.  Most
importantly, creating a pair of magnetic excitations for some $g\in G$,
moving them around an electric excitation $\alpha$, and annihilating them
again leaves us with a loop of $V_g$'s around $X_\alpha$, and thus yields
a non-trivial braiding phase equal to $\alpha(g)$, following
Eq.~\eqref{eq:X-irrep-phase}, illustrated in Fig.~\ref{fig:anyons}d.

For the RGFP model, where the tensor -- up to a basis transformation on
the physical system -- is nothing but a projector onto the 
invariant space of the symmetry \eqref{eq:G-injective}, these
anyon-like objects on
the entanglement level are mapped one-to-one to the physical level at the RGFP, 
that is to say, they can be created (in pairs), manipulated, and detected
by local physical operations (the operations just need to respect the
global $V_g$-symmetry). Thus, at the RGFP, these objects on the
entanglement level describe real anyons, that is, localized excitations
(quasi-particles) which are eigenstates of the Hamiltonian and have anyonic
statistics. These excitations are characterized by a group element $g$ and
an irreducible representation $\alpha$, and we will label them by $a\equiv
(g,\alpha)$, and its anti-particle by $\bar a\equiv (g^{-1},\bar\alpha)$
(here, $\bar\alpha$ denotes the complex conjugate).

\subsection{Behavior of anyonic operators vs.\ topological
order\label{sec:2:anyon-behavior}}

Do the objects which we have just constructed necessarily describe
topological excitations? They certainly possess the right properties at
the \emph{entanglement} level (we will call them ``virtual anyons''), but
does this necessarily mean they also describe proper \emph{physical
anyons}?  As just argued, at the RGFP this can easily be seen to be the case, due to
the unitary correspondence between the entanglement and physical degrees of
freedom on the invariant subspace \eqref{eq:G-injective} -- thus, the
anyonic operators at the entanglement level
can be created, manipulated, and detected by physical unitaries.  This
continues to holds as we move away from the RGFP -- we can 
understand this e.g.\ using quasi-adiabatic
evolution~\cite{hastings:quasi-adiabatic}, which effectively
evolves the tensors without affecting 
the entanglement 
symmetry \eqref{eq:G-injective}, and which will thus only dress the endpoints of
the strings (as in Fig.~\ref{fig:anyons}ab). In fact, this 
is precisely what underlies e.g.\ the
excitation ansatz for topological
excitations~\cite{vanderstraeten:excitation-ansatz-peps-forpeps,vanderstraeten:excitation-ansatz-peps-forall}. Without this dressing of the
endpoint, our virtual anyons might not be eigenstates of the Hamiltonian, but
they will regardless describe an excitation in the corresponding
topological sector (that is, a dispersing superposition of anyonic
excitation with
identical anyonic quantum number).

However, if we deform our tensors sufficiently strongly (e.g.\ towards a
product state), even while keeping the symmetry
\eqref{eq:G-injective}, topological order will eventually break down.
Yet, on the
entanglement level, the ``anyonic operators'' still possess the same
properties~\cite{schuch:topo-top}. This raises the question: How can we determine whether the 
virtual anyons in Fig.~\ref{fig:anyons}c do indeed
describe actual physical anyons? Or, equivalently, when is a system whose
wavefunction is described by tensors with a symmetry \eqref{eq:G-injective} 
truly topologically ordered?

As it turns out, whether the system is topologically ordered, and whether
the virtual anyons represent physical anyons, is precisely reflected 
in two properties, which we naturally demand from true anyonic 
excitations.  
\\[1ex]
\emph{\textbf{Properties of anyonic excitations:}}
To define the properties we require from anyonic excitations in the
topological phase, let us normalize our tensors such that the state is
normalized on the infinite plane,
\begin{equation}
\bra\Omega\Omega\rangle=1\ ,
\end{equation}
and let us denote by $\ket{\Psi_{a\bar a}(\ell)}$ the state with a pair
of ``virtual anyons'' $a$ and $\bar a$, Fig.~\ref{fig:anyons}c, placed at the entanglement
degrees of freedom at separation $\ell$. We require the following
properties from this state to describe a pair of physical anyons.
\begin{enumerate}
\item We need to be able to construct a well-defined, normalizable
wavefunction with individual anyons at arbitrary locations. This is
measured by the quantity
\begin{equation}
\label{eq:N_ell}
N_{a\bar a}(\ell) := 
    \bra{\Psi_{a\bar a}(\ell)}\Psi_{a\bar a}(\ell)\rangle\ .
\end{equation}
For well-defined anyonic excitations, we require $N_{a\bar a}(\ell)\to
K_a^2\ne 0$ as $\ell\to\infty$, such that $\ket{\Psi_{a\bar a}}$ is
normalizable for arbitrarily separated anyonic excitations.
\item 
Individual anyonic excitations must be orthogonal to the ground state,
as they are characterized by a non-trivial topological quantum number,
i.e., 
they live in a different (global) symmetry sector.
This is quantified by the overlap
\begin{equation}
\label{eq:F_ell}
F_{a\bar a}(\ell) := \big|\bra{\Psi_{a\bar a}(\ell)}\Omega\rangle\big|^2\ .
\end{equation}
We thus require that for non-trivial anyons $a$, $F_{a\bar a}(\ell)\to0$
as $\ell\to\infty$. (As
long as the anyons are close to each other, the total object $a\bar a$ has a trivial
topological quantum number and can thus have a non-zero overlap with the
ground state.) 
\end{enumerate}
Note that $0\le F_{a\bar a}(\ell)\le N_{a\bar a}(\ell)$, where the second
inequality is the Cauchy-Schwarz inequality.
It is thus natural to define a normalized quantity
\begin{equation}
\hat F_{a\bar a}(\ell):= F_{a\bar a}(\ell)/N_{a\bar a}(\ell)\le 1\ .
\end{equation}

In which way can the above two properties break down?  First, we can have that
for some anyon $a$, $N_{a\bar a}(\ell)\to 0$ as $\ell\to\infty$, that is,
we are unable to construct a well-defined state as we separate the anyons
$a$ and $\bar a$. In that case, we will say that the anyons $a$ and $\bar a$ 
are \emph{confined}. This implies that also $F_{a\bar a}(\ell)\to0$.
Second, we can have that for some anyon $a$, $F_{a\bar a}(\ell)\to C_a^2>0$
(and thus also $N_{a\bar a}(\ell)\to K_a^2>0$). In that case, the ``anyon'' $a$ is no
longer orthogonal to the ground state, that is, it is no longer
characterized by a distinct topological quantum number and thus  has
\emph{condensed} into the ground state.

We thus see that we for each ``virtual anyon'' $a$ constructed from the
entanglement symmetry and its antiparticle $\bar a$, we have three distinct possibilities:
\begin{enumerate}
\item \emph{\textbf{Free anyon:}} $N_{a\bar a}\to K_a^2>0$, $F_{a\bar a}\to
0$.
\item \emph{\textbf{Confined anyon:}} $N_{a\bar a}\to 0$.
\item \emph{\textbf{Condensed anyon:}} $\hat F_{a\bar a}\!\to\! \hat C_a^2\!>\!0$, 
$N_{a\bar a}\!\to\! K_a^2>0$.
\end{enumerate}
We call $\hat F_{a\bar a}$ the \emph{condensate fraction} and $N_{a\bar
a}$ the \emph{deconfinement fraction} for anyon $a$.

It turns out that these different behaviors can be used to identify the
different topological phases (including the trivial phase) compatible with
a given entanglement symmetry (\ref{eq:G-injective}) with symmetry group
$G$.  In fact, it has been shown to be in one-to-one correspondence to the
possible phases which can be obtained by the framework of anyon
condensation from the quantum double model $D(G)$.

\subsection{Anyonic operators as qualitative order
parameters\label{sec:2:opar-qual}}

As we have seen, the asymptotic behavior of $N_{a\bar a}(\ell)\to K_a^2$ and 
$F_{a\bar a}(\ell)\to C_a^2$ can serve as
order parameters which allow to \emph{distinguish} different topological
and trivial phases.  Let us now see how they can be related to
conventionally defined order parameters and string order
parameters~\cite{haegeman:shadows,duivenvoorden:anyon-condensation}. 
 This will not only be insightful on its
own right, but also provide us with guidance on how to use them as starting
points for the construction of \emph{quantitative} order parameters which
allow us to study universal behavior in the vicinity of topological phase
transitions.

To this end, let us consider the evaluation of $N_{a\bar a}(\ell)$ and
$F_{a\bar a}(\ell)$ in an iPEPS, where $a\equiv (g,\alpha)$.
There, both of these quantities take the form
\begin{equation}
\label{eq:anyonop-peps}
\raisebox{-2.5cm}{\includegraphics[scale=0.88]{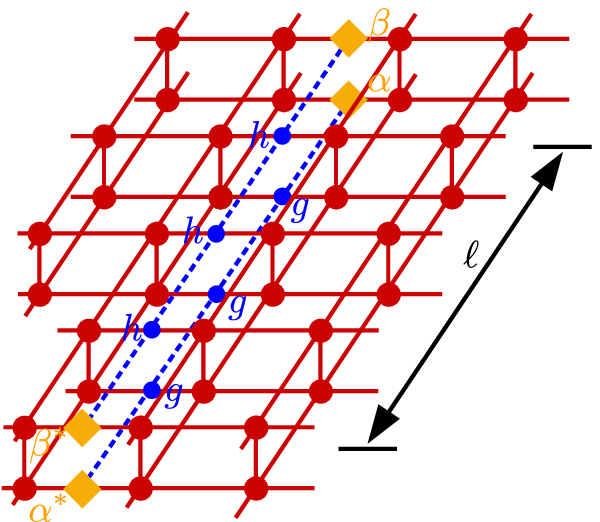}}\ ,
\end{equation}
that is, they are string-like operators which are evaluated along a cut in
the (infinite) PEPS. 
 Specifically, for $N_{a\bar a}(\ell)$, $h=g$ and $\beta=\alpha$, while
for $F_{a\bar a}(\ell)$, $h=\gid$ (the identity element of $G$)  and $\beta=1$.
In order to evaluate those quantities, one proceeds 
as follows: Denote by
\begin{equation}
\label{eq:transfer-operator}
\raisebox{-1.1cm}{\includegraphics{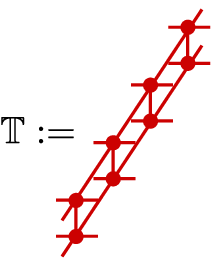}}
\end{equation}
the transfer operator, that is, one column of Eq.~\eqref{eq:anyonop-peps}.
Then, determine the left and right fixed points $\sigma_L$ and $\sigma_R$
of $\mathbb T$. Numerically, this is done by approximating $\sigma_L$ and
$\sigma_R$ with iMPS of bond dimension $\chi_L$ and $\chi_R$ (with tensors
$M_L$ and $M_R$); this is justified by the
fact that in gapped phases, correlations decay exponentially and thus iMPS
provide a good approximation (the quality of which can be assessed by
increasing
$\chi$)~\cite{brandao:exp-corr-arealaw,verstraete:faithfully,haegeman:medley}.  One thus finds that the evaluation of the anyon
behavior reduces to evaluating the one-dimensional object
\begin{equation}
\label{eq:anyon-opar-at-bnd}
\raisebox{-.9cm}{\includegraphics{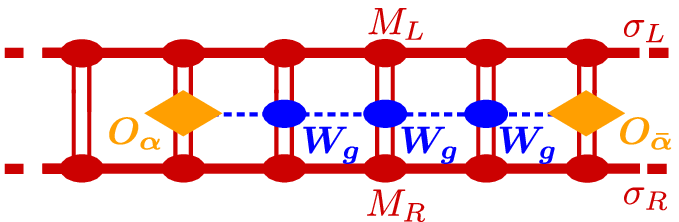}}\quad,
\end{equation}
where we have defined the double-layer symmetry operators $\bm{W_g} :=
V_g\otimes \bar V_h$
with $\bm g=(g,h)$, and double-layer operators $\bm{O_\alpha}$ which
transform as the irrep $\bm\alpha(\bm g):=\alpha(g)\bar\beta(h)$ of
$\bm G:=G\times G$, $\bm{W_g}\bm{O_\alpha}\bm{W_g}^\dagger =
\bm{\alpha}(\bm{g})\bm{O_\alpha}$.

The fact that the fixed points of $\mathbb T$ are well approximated by MPS
is very resemblant of ground states of local Hamiltonians. In turn, the
fact that those ground states are well described by MPS
is constitutive of their physics and the types of order they
exhibit~\cite{pollmann:symprot-1d,chen:1d-phases-rg,schuch:mps-phases}. 
Thus, it is suggestive to analyze the above expression from the
perspective of $\sigma_{L,R}$ being ground states of an ``effective
Hamiltonian'' defined through $\mathbb T=e^{-\mathbb H}$. This Hamiltonian (just as
$\mathbb T$) possesses a symmetry 
\begin{equation}
\label{eq:top-sym-GG}
[\mathbb H,\bm{W_g}^{\otimes N}]=0\ 
\end{equation}
which it inherits from Eq.~(\ref{eq:G-injective}).  
Viewed from this
angle, we see (and will discuss further in a moment) that the expressions in Eq.~\eqref{eq:anyon-opar-at-bnd} can be
understood as \emph{(string) order parameters} for the symmetry $G\times
G$, Eq.~\eqref{eq:top-sym-GG}, measured in the ``ground state'' of
$\mathbb H$, i.e., the fixed point of $\mathbb T$.  Differently speaking,
they represent order parameters at the boundary, that is, in the
entanglement spectrum. 
   Note that $\mathbb T$ (and thus $\mathbb H$) is not
hermitian, and thus has different left and right fixed points, which leads
to additional subtleties when making analogies to the Hamiltonian case.

To better understand the structure behind these operators, 
let us first discuss
conventional order parameters from a bird's eye perspective, using the
minimum information possible.  This will
allow us to reason by analogy in the discussion of topological order
parameters, but at the same time also help us to flesh out those aspects
where the current situation is fundamentally different and poses novel
challenges.  As guidance, we will consider models $H$ with a $\mathbb Z_2$
symmetry $[H,Z^{\otimes N}]=0$ with
\begin{equation}
\label{eq:general-z2-sym}
Z=\begin{pmatrix}\openone_{D_e}\\&-\openone_{D_o}\end{pmatrix}
\end{equation}
with some degeneracy $D_e$ and $D_o$ of the two irreps.  
As a specific example, we will keep
returning to the $(1+1)$D transverse field
Ising model 
\begin{equation}
\label{eq:ising-model}
H = \sum X_i X_{i+1} + h \sum Z_i\ ,
\end{equation}
(with $X$, $Z$ the Pauli matrices, i.e., $D_e=D_o=1$),
but we will also find that the case where $D_e,D_o>1$
holds additional challenges. The following considerations 
will similarly also hold for more general symmetry groups
$G$ with 
representations $W_g$, $g\in G$.  (We limit the use of boldface 
notation to when interested specifically in the double-layer structure of the
PEPS.)

A key point in the
symmetry-breaking paradigm of studying phases is that a priori, all we are
supposed to use is the symmetry itself, and not additional properties of
the concrete $H$ given. This is particularly important in the situation
at hand, where for the transfer operator $\mathbb T$ and the underlying
Hamiltonian $\mathbb H$, all we know is indeed the symmetry 
\eqref{eq:top-sym-GG}. (Recall that we consider PEPS tensors obtained from
a full variational optimization where solely the symmetry is imposed.)

For the Ising model above, one would usually choose $X$ as the order parameter.
However, this choice is not at all unique: Based solely on the symmetry, any
other operator $O$ with $ZOZ^\dagger=-Z$ (that is, $O=\cos\theta\,X +
e^{i\phi}\sin\theta\, Y$) will serve the same purpose, namely to be zero in
the disordered (symmetric) phase due to symmetry reasons, and generically
non-zero in the ordered (symmetry-broken) phase
except for fine-tuned choices of $\theta$ and $\phi$.
A dual way of seeing this is to
notice that the Ising Hamiltonian \eqref{eq:ising-model} can be
arbitrarily rotated in the XY plane while preserving the $\mathbb Z_2$
symmetry. The same principle holds for more general symmetries and/or
other representations: All that matters for an order parameter is that it
transforms as a non-trivial irreducible representation of the symmetry
group, $W_g O_\alpha W_g^\dagger = \alpha(g) O_\alpha$. Indeed, there
is not even the need to restrict to single-site operators -- any operator
acting on a finite range, such as $O=X\otimes X\otimes X$, will share
those properties; this point will become relevant later on.

Order parameters are directly tied to correlation functions: Given an
order parameter $O\equiv O_\alpha$ which transforms as an irrep $\alpha$,
we can consider the correlation function $\langle O_i O_j^\dagger\rangle$
between $O$ at
position $i$ and $O^\dagger$ (transforming as $\bar\alpha$) at $j$, which will go
to zero in the disordered phase and to a non-zero
constant in the ordered phase, namely $|\langle O\rangle|^2$ evaluated in
a symmetry broken state.  $\langle O_i O_j^\dagger\rangle$ has the
advantage that unlike $\langle O\rangle$, it transforms trivially under
the symmetry and thus does not depend on the
state in which it is evaluated (this
is used e.g.\ in Quantum Monte Carlo simulations). Note that at the same
time, in the disordered phase $\langle O_i O_j^\dagger\rangle$ will decay
exponentially to zero (as long as it is a gapped phase), and thus any
order parameter $O$ also defines a length scale at the other side of the phase
transition.

Comparing this discussion with Eq.~\eqref{eq:anyon-opar-at-bnd}, we see
that $\langle O_i O_j^\dagger\rangle$ 
is indeed one of the objects which appear there, namely for
$g=g'=\mathrm{id}$.   However, there are also other quantities appearing in
Eq.~\eqref{eq:anyon-opar-at-bnd}, such as the expectation value of a
string of symmetry operations, $\langle W_g\otimes\cdots\otimes
W_g\rangle$.  In the Ising model, this would amount to measuring the
expectation value of a string 
$\langle Z_i\otimes\cdots\otimes Z_j\rangle$. This operator has a natural
interpretation: In the symmetry broken phase, it flips the spins in a
region and thereby creates a pair of domain walls.  Thus, 
after applying $Z_i\otimes \cdots\otimes Z_j$,
the spins
between $i$ and $j$ are magnetized in the opposite direction, and 
$\langle Z_i\otimes\cdots\otimes Z_j\rangle\to0$ as $|i-j|\to\infty$. On
the other hand, in the disordered phase, this only creates local defects
at the endpoint, and thus 
$\langle Z_i\otimes\cdots\otimes Z_j\rangle\to\mathrm{const.}$; this
constant can be seen as an order parameter corresponding to a
semi-infinite string of $Z_i$'s (a soliton).  Note that under the self-duality of the
Ising model, such a semi-infinite string of $Z$'s is exchanged with an $X$
at its endpoint, that is, it is the order parameter for the dual model,
which is non-zero in the \emph{disordered} phase (sometimes termed a
``disorder parameter'').  

In fact, this is a special case of a \emph{string order parameter}, that is, a
correlation function of the form $\langle O_i \otimes
W_g\otimes\cdots\otimes W_g\otimes O^\dagger_j\rangle$, where $O$ transforms as
an irrep $\alpha$ of the symmetry group.   String
order parameters can be used to characterize 
both conventional (symmetry breaking) \emph{and}
symmetry protected (SPT)
phases in 1D, and their pattern is in one-to-one correspondence to the
different SPT phases (specifically, 
the non-zero string-order parameters satisfy
$\alpha(h) = \omega(g,h)/\omega(h,g)$, where $\omega$ is the $2$-cocycle
characterizing the SPT
phase)~\cite{pollmann:spt-detection-1d,duivenvoorden:anyon-condensation}.
In fact, this is exactly what happens above in
Eq.~\eqref{eq:anyon-opar-at-bnd}: The behavior of anyons is in one-to-one
correspondence to string order parameters at the boundary under the
$G\times G$ symmetry, Eq.~\eqref{eq:top-sym-GG}; indeed, it has been shown
that the possible ways in which anyons can condense and confine is in
exact correspondence to the possible SPT phases under the symmetry group
$G\times G$, if one additionally takes into account the constraints from
positivity of
$\sigma_L,\sigma_R\ge0$~\cite{duivenvoorden:anyon-condensation}. 

In the following, we will use the terminology ``order parameter'' to refer
to both ``conventional'' order parameters and string order parameters
equally.

\subsection{Anyonic operators as quantitative order
parameters\label{sec:anyopar-quantitative-and-gauge}}

Up to now, we have discussed the interpretation of anyonic operators as
order parameters 
for the \emph{detection and disambiguation} of different phases 
under the topological symmetry $\bm{W_g}=V_g\otimes \bar V_h$ of the
transfer operator.  But order parameters can also be used to
\emph{quantitatively} study transitions between different phases and investigate
their universal behavior.  In the following, we will discuss whether and
how we can use anyonic operators to the same end, that is, for a
\emph{quantitative} study of topological phase transitions. However, as we will
see, the situation has a number of additional subtleties as opposed to the
conventional application of order parameters.  Those subtleties do not
a priori arise from fundamental differences between topological vs.\
conventional phase transitions. Rather, they stem from the fact that for
PEPS obtained from a \emph{variational} optimization in which \emph{only}
the topological symmetry \eqref{eq:G-injective} has been imposed -- which is
what what we focus on in this work -- \emph{all} we know for sure about the transfer
matrix $\mathbb T$ and thus about the effective Hamiltonian $\mathbb H$ is
that it possesses that very same symmetry, Eq.~\eqref{eq:top-sym-GG}.
This is rather different from physical Hamiltonians or engineered
variational ``toy models'' (as e.g.\ in
Refs.~\cite{verstraete:comp-power-of-peps,iqbal:z4-phasetrans,iqbal:rvb-perturb,xu:z2-phase-transitions-tn,xu:fib-phase-trans-tn,vanderstraeten:peps-perturbations,schotte:fibonacci-perturbative-peps}),
where we have a smooth dependence of $H(\lambda)$ or $\mathbb H(\lambda)$ on the external
parameter.

How is this smooth dependence relevant?  Let us illustrate this with the
Ising model, or generally models with a $\mathbb Z_2$ symmetry
\eqref{eq:general-z2-sym}.  If the Hamiltonian $H(\lambda)$ depends
smoothly on the parameter $\lambda$, such as in the Ising model, we can
choose any fixed local operator which anticommutes with the symmetry as
our order parameter, such as $X$.  However, let us now consider a
``scrambled'' version of the Ising model,
\begin{equation}
\label{eq:random-ising}
H_s(\lambda) = R(\lambda)^{\otimes N}\, H(\lambda) 
 (R(\lambda)^\dagger)^{\otimes N}\ ,
\end{equation}
where for each value of $\lambda$, we apply a \emph{random gauge}
$R(\lambda)$ which commutes with the symmetry; that is, $R(\lambda) =
\exp(i\theta(\lambda)Z/2)$ is a rotation about the $z$ axis by
an angle $\theta(\lambda)$ which is chosen at random \emph{separately}
for each value of $\lambda$.\footnote{In the light of the non-hermiticity of
$\mathbb T$ and $\mathbb H$, and the non-unitarity of the gauge
\eqref{eq:peps-gauge-dof}, we also allow for non-unitary $R$, corresponding here to
complex values of $\theta$.}
While this seems contrived for an actual Hamiltonian, this is exactly
the situation we must expect to face in our simulation: The variationally
optimized tensor can come in a random basis -- that is, with a random
gauge choice $Q$ and $R$ in Eq.~\eqref{eq:peps-gauge-dof} -- for each value of the
parameter $\lambda$ independently, and the only property we are guaranteed
is that it possesses the symmetry \eqref{eq:G-injective}, and thus the
gauge commutes with the symmetry, $[Q,V_g]=[R,V_g]=0$. 

Clearly, picking a fixed order parameter such as $X$ will not work for the
randomly rotated Hamiltonian \eqref{eq:random-ising}, as it would yield
the ``normal'' Ising order parameter $\langle X\rangle$ modulated with a 
random amplitude $\cos(\theta(\lambda))$, and thus be random itself.  A
way around could be to \emph{maximize} the value of the order parameter
over all single-site operators $O$ with $ZOZ^\dagger=-O$ (or even all
$k$-site operators for some fixed $k$). However, while this approach will likely
work well in the scenario above, it is not a viable approach in the case
of anyonic operators in PEPS. The reason
is that in a PEPS, local objects on the entanglement level, or e.g.\ a
modified tensor, can affect the
PEPS on a length scale of the order of the correlation length (and in
principle even beyond, at the cost of singular behavior), which is
precisely the reason why e.g.\ PEPS excitation ansatzes work 
even though they only change a single
tensor~\cite{haegeman:mps-ansatz-excitations,vanderstraeten:excitation-ansatz-peps-forall}.
In our case, however, this would amount to allowing 
optimization over $O$ which are supported on a region on the order of the
correlation length.  In that case, it is easy to see that this approach is
bound to fail:  Specifically, in the case of the (non-gauge-scrambled) Ising
model, we can take the RGFP order parameter $X$ and
quasi-adiabatically~\cite{hastings:quasi-adiabatic}
continue it with $\lambda$, to obtain an effective order parameter
$X(\lambda)$ with expectation value $\langle X(\lambda)\rangle_\lambda
\equiv \langle X\rangle_{\lambda=0}=1$ all the way down to the phase
transition, and where $X(\lambda)$ is approximately supported on a region of the order of
the correlation length.  We thus see that an order parameter which is
optimized over such a growing region will yield the value $1$ all the way
down to the phase transition, and thus not allow to make quantitative
statements about the nature of the transitions.\footnote{We have checked
this for the model presented in Sec.~\ref{sec:tc-field} and indeed found
that optimizing the order parameter (at fixed operator norm) such as to
maximize its expectation value gives a curve which approaches a step
function as the bond dimension $D$ grows.}

We thus require another way to obtain well-defined order parameters. A
natural approach would be to choose order parameters
which are gauge-invariant, 
that is, order parameters which are constructed such as to be invariant under a
random gauge choice.  For a local order parameter alone, however, this is not
possible, since
$ZOZ^\dagger=-O$ implies
$O=\left(\begin{smallmatrix}0&a\\b&0\end{smallmatrix}\right)$, which
transforms under
$R(\lambda)=\left(\begin{smallmatrix}c_0\\&c_1\end{smallmatrix}\right)$
as
\begin{equation}
R(\lambda) O R(\lambda)^{-1} = 
\begin{pmatrix} 0 &  c_0 a c_1^{-1}\\ c_1 b c_0^{-1} & 0\end{pmatrix}\ ,
\end{equation}
which will never be gauge invariant, independent of the choice of $a$ and
$b$.
However, there still \emph{is} a way to measure the order parameter in a gauge
invariant way:  To this end, define a pair of order parameters $O=
\left(\begin{smallmatrix}0&1\\0&0\end{smallmatrix}\right)$ and $O^\dagger=
\left(\begin{smallmatrix}0&0\\1&0\end{smallmatrix}\right)$, and measure
$\langle O_i O^\dagger_j\rangle$ for $|i-j|\to\infty$.
Let us now see what happens to this object under a gauge transformation
$R$: $O$ acquires a factor $c_0c_1^{-1}$, while $O^\dagger$ acquires
$c_1c_0^{-1}$. In the correlator $\langle O_i O_j^\dagger\rangle$, the
gauge therefore cancels, and we obtain a well-defined, gauge-invariant
quantity.  Thus, we see that we can obtain a gauge-invariant order
parameter by combining \emph{pairs} of order parameters for which the
gauges cancel and measuring the corresponding correlator for $\ell\to\infty$.
(We can then e.g.\ assign the square root of the correlation to each of
the order parameters.)
The same idea also works for general abelian symmetries, as long as all
irreps are non-degenerate: In that case, the symmetry $[O_\alpha,W_g]=0$
limits the non-zero entries of $O_\alpha$ to be $(O_\alpha)_{i,i+\alpha}$,
which under a gauge $R=\mathrm{diag}(c_0,c_1,\dots)$ acquire a prefactor
$c_i c_{i+\alpha}^{-1}$. Thus, by choosing $O_\alpha=\delta_{i,i+\alpha}$
for an arbitrary $i$, $O_\alpha$ and $O_\alpha^\dagger$ acquire opposite
prefactors and thus yield again gauge-invariant correlators.

So does this allow us to define a gauge-independent order parameter?
Unfortunately, this is only partly the case: As soon as we have symmetries
with degenerate irrep spaces, such as in \eqref{eq:general-z2-sym}, any
generalized gauge transformation of the form
\begin{equation}
R=\begin{pmatrix} R_0 \\ & R_1\end{pmatrix}
\end{equation}
is admissible, under which an order parameter
$O = \left(\begin{smallmatrix} & A\\B\end{smallmatrix}\right)$
transforms as 
\begin{equation}
ROR^{-1} = \begin{pmatrix} & R_0AR_1^{-1}\\R_1BR_0^{-1}\end{pmatrix}\ .
\end{equation}
In this case, no gauge invariant choice can be made, since $\langle
ROR^{-1}\rangle$ is evaluated in the reduced density matrix at that site,
about which we do not have any additional information a priori.  In
particular, the dependence of the two endpoints on $G$ will not cancel out,
even if we set $A$ or $B$ to $0$, respectively; nor does a special choice
like $A=B=\openone$ help (as it leaves us e.g.\ with $R_0R_1^{-1}$). In that
case, we must rely on a way of \emph{fixing} a
smooth gauge for the Hamiltonian $H$ (or $\mathbb H$); we will 
explain the concrete recipe in Section~\ref{sec:subsec:recipe}.

A special case is given by order parameters which only involve
semi-infinite strings of symmetry operators $\cdots\otimes W_g\otimes
W_g\otimes \openone\dots$ (in the context of topological order, these measure
flux condensation and deconfinement); in the case of the Ising model, we saw that
they created domain walls in the
symmetry broken phase and were dual to the usual order parameters.  These
order parameters have the feature that they \emph{are} gauge invariant,
since any gauge $R$ must satisfy $[R,W_g]=0$ -- they thus have a well-defined value and can be
measured without involving any additional gauge fixing.  
Note, however, that this only holds for string order parameters with a
trivial endpoint.  In case the model has dualities between those ``pure''
string order parameters and other order parameters, we can additionally
use these dualities to measure further order parameters directly in a gauge
invariant way.

\subsection{A practical summary: How to compute anyonic order parameters
in iPEPS\label{sec:subsec:recipe}}

In the following, we summarize our finding in the form of a practical
recipe: How do I compute anyonic order parameters for a model Hamiltonian
using iPEPS? Again, we will
focus on Abelian symmetry groups $G$. Our starting point is always a
physical Hamiltonian model $\mathcal H\equiv \mathcal H(\lambda)$, for
which we optimize the energy variationally.  

In the first step, we need to define the
overall setting:  The way in which the symmetries are imposed on
the tensors,  which is the same for all values of the parameter~$\lambda$.

\vspace*{2ex}
\noindent\emph{\textbf{I. Define symmetries:}}
\begin{enumerate}
\item 
Pick the appropriate symmetry group $G$ for the system at hand, together
with a representation $V_g=\bigoplus_\alpha
\alpha(g)\otimes\openone_{d_\alpha}$ with irreps $\alpha(g)$. (Note that
we work in a basis where $V_g$ is diagonal.)
\item
Define ``endpoint operators'' $X_{\alpha,\gamma}$ (representing charges
$\alpha$,  cf.\ Fig.~\ref{fig:anyons}c) as
\begin{equation}
\label{eq:recipe:endpoint-ops}
X_{\alpha,\gamma} = \delta_{\gamma+\alpha,\gamma}\otimes M_{\alpha,\gamma}
\end{equation}
for some $M$ -- that is, $X_{\alpha,i}$ only has non-zero elements in row
and column with irrep $\gamma+\alpha$ and $\gamma$,
respectively.\footnote{The irreps form an additive group which we denote
by $+$, even though we also choose to denote the inverse of $\alpha$ by
$\bar\alpha$.} We
choose $M_{\alpha,\gamma}=\openone$ (this requires that the two
irreps $\gamma+\alpha$ and $\gamma$ have the same dimension), 
other choices are discussed in Sec.~\ref{sec:discussion-endpoints}.
\end{enumerate}

\vspace{1em}\noindent
Now, we can perform a PEPS optimization and compute order parameters for
each $\lambda$ and $\mathcal H\equiv \mathcal H(\lambda)$; we will suppress the
$\lambda$-dependence in the following.

\vspace*{2ex}
\noindent\emph{\textbf{II. Compute order parameters:}}
\begin{enumerate}
\item Optimize the iPEPS tensor $A$ subject to the symmetry 
\begin{equation}
\label{eq:ginj-recipe}
\raisebox{-0.7cm}{\includegraphics{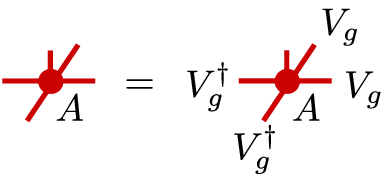}}\quad,
\end{equation}
such as
to minimize the energy with respect to the Hamiltonian $\mathcal H\equiv
H(\lambda)$.
This can be accomplished, e.g., by using a gradient method and projecting
the gradient back to the symmetric space \eqref{eq:ginj-recipe}, or using
a tangent-space method on the manifold of symmetric PEPS.\footnote{As usual in PEPS
optimizations, the correct choice of the initial tensor can be relevant.
Experience shows that one should choose an initial tensor in the
topological phase. Moreover, changing tensors adiabatically in $\lambda$
can give more stable results. See Sec.~\ref{sec:tc-field}  for further
discussion.}
\item 
\label{item:recipe:gaugefixing}
Consider the tensor
\begin{equation}
\label{eq:recipe:Ch}
\raisebox{-0.8cm}{\includegraphics{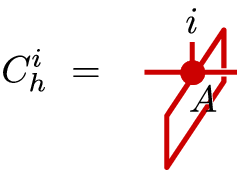}}\quad .
\end{equation}
with $i$ the physical index.
This is an MPS tensor with symmetry $V_g$, $C_h^i=V_g^\dagger C_h^i V_g$.
Apply the MPS gauge fixing described in
part IIa
below. This yields a gauged tensor $\tilde C_h$ and a gauge $Q=\bigoplus
Q_\alpha$ which commutes with the symmetry,
\begin{equation}
\label{eq:recipe:Ch-tilde}
\raisebox{-0.8cm}{\includegraphics{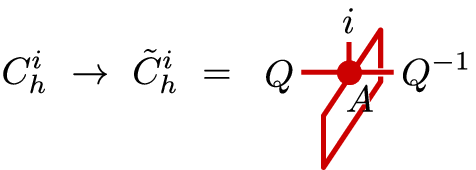}}\quad .
\end{equation}
Similarly, consider the tensor $C_v$ obtained from closing the indices
horizontally and perform the analogous gauge fixing, yielding a gauge
$R=\bigoplus R_\alpha$:
\begin{equation}
\label{eq:recipe:Cv}
\raisebox{-0.6cm}{\includegraphics{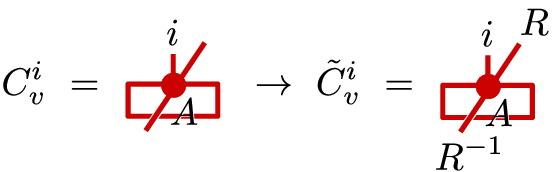}}.
\end{equation}
The gauge-fixed PEPS tensor $\tilde A$ is then obtained as 
\begin{equation}
\raisebox{-0.7cm}{\includegraphics{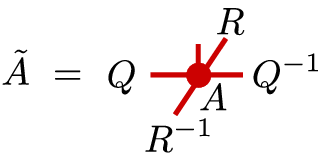}}\quad .
\end{equation}
\item 
Compute the PEPS environment $\rho(g,h)$ for a single site from the
gauge-fixed tensor $\tilde A$, with a
semi-infinite string of group actions $V_g\otimes \bar{V}_h\equiv
\bm{W_g}$ attached (including the identity operator $g,h=\gid$):
\begin{equation}
\hspace*{1em}\raisebox{-1.4cm}{\includegraphics{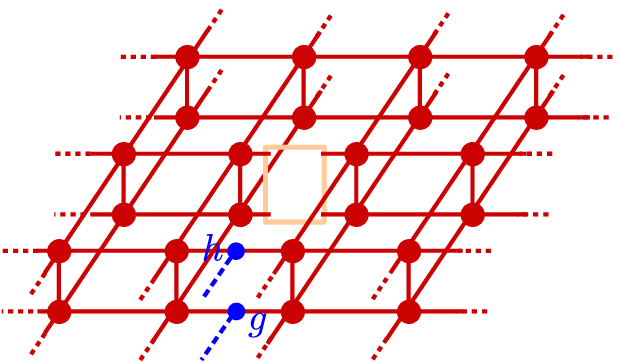}}\ .
\end{equation}
(The four indices of $\rho(g,h)$ are marked by the orange box.)
For instance, this can be done by computing the iMPS fixed point of the transfer
operator from left and right, with tensors $M_L$ and $M_R$, cf.\
Eq.~\eqref{eq:anyon-opar-at-bnd},
and then contracting the ``channel operator'' with a string on one side,
\begin{equation}
\hspace*{2em}\raisebox{-1.0cm}{\includegraphics{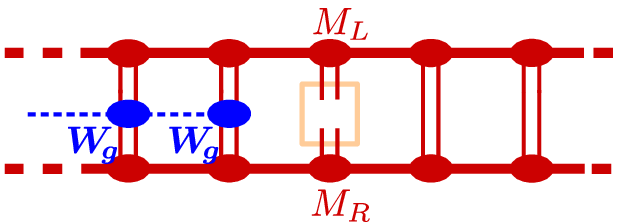}}\ ,
\end{equation}
where $\bm{W_g}=V_g\otimes \bar V_h$.
Alternatively, one can e.g.\ also use a CTM-based method.
\item Define the normalizations
\begin{align}
\label{eq:recipe:N_g_a_g}
\mathcal N(g,\alpha,\gamma) &= 
    \mathrm{tr}[\rho(g,g)\:
	X_{\alpha,\gamma}\otimes \bar X_{\alpha,\gamma}]
\\
\mathcal N_\mathrm{vac} &= 
    \mathrm{tr}[\rho(\mathrm{id},\mathrm{id})\:
	X_\mathrm{vac}\otimes \bar X_{\mathrm{vac}}]
\end{align}
and the overlaps
\begin{equation}
\label{eq:recipe:O_g_a_g}
\mathcal O(g,\alpha,\gamma) =
    \mathrm{tr}[\rho(g,\mathrm{id})\:
	X_{\alpha,\gamma}\otimes \bar X_\mathrm{vac}]\ ,
\end{equation}
where $X_\mathrm{vac}=\openone=\sum X_{0,\gamma}$.
\item 
The condensate fraction of anyon $a=(g,\alpha)$ and its anti-particle
$\bar a=(g^{-1},\bar\alpha)$ is obtained as 
\begin{equation}
\label{eq:condfrac-recipe}
\hat C_{a,\gamma} =  \frac{
\sqrt{\mathcal O(g,\alpha,\gamma)\:\mathcal O(g^{-1},\bar\alpha,\gamma')}
}{
\sqrt{\mathcal N(g,\alpha,\gamma)\:\mathcal N(g^{-1},\bar\alpha,\gamma')}
\;
\mathcal N_\mathrm{vac}
}
\end{equation}
with $\gamma'=\gamma+\bar\alpha$, which ensures that $\hat C_{a,\gamma}$ is gauge-invariant.
Note that $\hat C_a\equiv \hat C_{a,\gamma}$ can depend on the choice of
$\gamma$,
but we expect all of them
to exhibit the same universal behavior.

\item 
The deconfinement fraction is obtained as 
\begin{equation}
\label{eq:conffrac-recipe}
K_{a,\gamma} = \frac{\sqrt{\mathcal N_{g,\alpha,\gamma}\:
    \mathcal N_{g^{-1},\bar\alpha,\gamma'}}}{\mathcal N_\mathrm{vac}}
\end{equation}
with $\gamma'$ as before. Again, $K_{a,\gamma}$ can depend on $\gamma$ and the
choice of vacuum, but with the same universal behavior.
\end{enumerate}

\vspace*{2ex}
\noindent\emph{\textbf{IIa. Gauge fixing:}}
Let us now describe the gauge fixing procedure used in step
II.\ref{item:recipe:gaugefixing} above for the tensors in
Eqs.~\eqref{eq:recipe:Ch} and \eqref{eq:recipe:Cv}. 

In either case, we are given an MPS tensor $C\equiv C^i$ with
$C^i=V_g^\dagger C^i V_g$, that is, the $C^i$ are diagonal in the irrep
basis of $V_g$: $C^i = \bigoplus_\alpha C^i_\alpha$. The key point in the
following is that the gauge fixing must uniquely fix \emph{all} gauge
degrees of freedom.

The following gauge fixing procedure is then carried out individually for
each irrep sector $C^i_\alpha\equiv B^i$.

\begin{enumerate}
\item Fix the right fixed point (i.e., the leading right eigenvector) of the transfer matrix $\mathbb E = \sum_i
\bar B^i\otimes B^i$ to be the identity. 
To this end, compute the leading right eigenvector $\rho\ge0$ of $\mathbb
E$ and replace $B^i$ by $B^i_r = \rho^{1/2} B^i \rho^{-1/2}$.
\item Fix the left fixed point of $\mathbb E_r=\sum_i \bar B^i_r\otimes
B^i_r$ to be diagonal with decreasing entries. To this end, compute the leading left eigenvector
$\sigma\ge0$ of $\mathbb E_r$, diagonalize it as $\sigma=U\Lambda U^\dagger$
with $\Lambda$ diagonal and decreasing and $U$ unitary, and let $B^i_{rl}=U^\dagger B^i_r
U$. (Note that this has to be done consistently with the index ordering
chosen for $\sigma$.)
\item There is a remaining degree of freedom: Both the left and right
fixed point remain invariant if we conjugate $B_{rl}^i$ with a diagonal
phase matrix $S$. To fix this degree of freedom, choose the diagonal of
$S$ equal to the phase of the first row of $B^1_{rl}$, and set the first entry
of $S=1$. Then, $\tilde B^i=SB^i_{rl}S^{-1}$ has positive entries on the
first row (except possibly the diagonal entry).  This uniquely fixes the
remaining phase degrees of freedom up to an irrelevant global phase.
\item The overall gauge transformation $O$, $B^i\to\tilde B^i = O B^i
O^{-1}$, is then given by 
\begin{equation}
O = S\,U^\dagger\,\rho^{1/2}\ .
\end{equation}
Importantly, $O$ is uniquely determined: $\rho$ is uniquely determined
(with eigenvalue decomposition $\rho=VDV^\dagger$), and $U^\dagger$ is
determined up to left-multiplication by a diagonal phase matrix, which is
subsequently fixed by $S$. Thus, $SU^\dagger
\rho^{1/2}=(SU^\dagger V)\,D\,V^\dagger$ uniquely fixed all free parameters in
the singular value decomposition of $O$.
\end{enumerate}
The steps above give a gauge fixing $O\equiv Q_\alpha$ for each irrep
block $\alpha$, $B^i\equiv C^i_\alpha$. The overall gauge fixing for
$C^i$, $C^i\to\tilde C^i=QC^iQ^{-1}$, is then given by $Q=\bigoplus
Q_\alpha$. Note, however, that this does not fix the relative weight of
different irrep blocks; this is taken care of by considering 
order parameters which are invariant under this gauge, namely pairs of
endpoints where the respective gauge degrees of freedom cancel out.

Note that the gauge fixing procedure is highly non-unique, and different
procedures can be used; however, we found that they do not affect the
universal behavior observed. For instance, one could replace the choice of
one identity and one diagonal fixed point by a gauge where both fixed
points are chosen to be equal.  Maybe more importantly, the phase fixing
is rather arbitrary, and in certain situations might have to be replaced
by a different procedure, such as when the entries used to fix $S$ are
very small, in which case on could e.g.\ pick a different combination of
matrix elements.

\vspace*{2ex}
\noindent\emph{\textbf{III. Anyon lengths (mass gaps) and confinement
length:}}
In addition to order parameters, we can also extract  anyon masses $m_a$,
that is, the correlation length $\xi_a=1/m_a$ associated to a given anyon,
for free anyons.  Specifically, $\xi_a$ is the correlation length
associated to the exponential decay of $F_{a\bar a}(\ell)\sim
e^{-\ell/\xi_a}$, Eq.~\eqref{eq:F_ell}, that is, the overlap of the PEPS
with anyons $a$ and $\bar a$ placed at distance  $\ell$ with the vacuum.
On the other hand, for confined anyons, a ``confinement length'' $\xi_{a\bar a}^K$
can be extracted -- this is the length
scale associated to the exponential decay of $N_{a\bar
a}(\ell)\sim e^{-\ell/\xi^K_{a\bar a}}$.
To extract these lengths, proceed as follows:
\begin{enumerate}
\item Define
\begin{equation}
\mathbb E^{g',\alpha'}_{g,\alpha} =\ 
\raisebox{-1cm}{\includegraphics{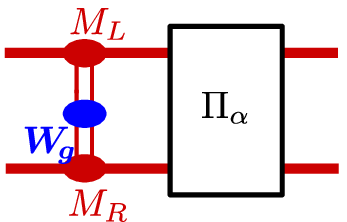}}\quad ,
\end{equation}
where $\bm{W_g}=V_g\otimes {\bar V}_{g'}$, $\bm g=(g,g')$, and 
$\Pi_{\bm\alpha} = \tfrac{1}{|G|^2}\sum_{h,h'}
\alpha(h)\bar{\alpha}'(h')Y_{\bm{h}}$ 
is the projection onto irrep sector
$\bm\alpha=(\alpha,\alpha')$. Here, $Y_{\bm{h}}$ is
the rotation on the ``virtual virtual'' indices of $M_R$ corresponding to
$\bm{W_h}$, $\bm h=(h,h')$, i.e.
\begin{equation}
\hspace*{1em}\raisebox{-.8cm}{\includegraphics{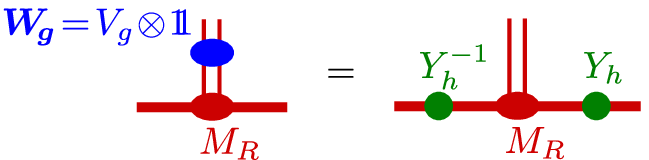}}\quad .
\end{equation}
(It can e.g.\ be computed by comparing the fixed point $\rho$ of the transfer
matrices $\sum(\bar M_R)_{ij}\otimes (M_R)_{ij}$ and $\rho_{\bm{W_h}}$ of the
dressed transfer matrix $\sum(\bar
M_R)_{i'j}(\bm{W_h})_{i'i}\otimes (M_R)_{ij}$, which are related as
$\rho_{\bm{W_h}} =
\rho Y_{\bm{h}}$; this can be facilitated by bringing 
$M_L$ into  canonical form such that $\rho=\openone$, which also yields a unitary
$Y_{\bm h}$~\cite{pollmann:spt-detection-1d,iqbal:z4-phasetrans}.)
\item Let $\lambda_1(X)$ and $\lambda_2(X)$ denote the two eigenvalues
of $X$ with largest magnitude. Then, the mass gap in the topologically
trivial sector is 
\begin{equation}
\hspace*{2em}
m_\mathrm{vac} = 1/\xi_\mathrm{vac} = 
-\log\big|\lambda_2(\mathbb E^{\gid,1}_{\gid,1})/\lambda_1(\mathbb
E^{\gid,1}_{\gid,1})\big|\ ,
\end{equation}
and the mass gap of a non-trivial anyon $a=(g,\alpha)\ne(\gid,1)$ is given by
\begin{equation}
\hspace*{2em}
m_a = 1/\xi_a = 
-\log\big|\lambda_1(\mathbb E^{\gid,1}_{g,\alpha})/\lambda_1(\mathbb
E^{\gid,1}_{\gid,1})\big|\ .
\end{equation}
Finally, the confinement length is given by
\begin{equation}
\hspace*{2em}
\xi^K_{a\bar a}=
-1/\log\big|\lambda_1(\mathbb E^{g,\alpha}_{g,\alpha})/\lambda_1(\mathbb
E^{\gid,1}_{\gid,1})\big|\ .
\end{equation}
\end{enumerate}

\section{Toric Code in a magnetic field\label{sec:tc-field}}

\subsection{Model and tensor network representation}

\begin{figure}[b]
\includegraphics[width=\columnwidth]{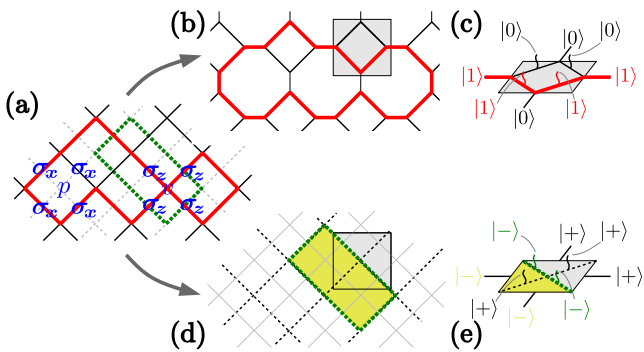}
\caption{The two dual PEPS representations of the Toric Code ground state.
\textbf{(a)} The Toric Code can be seen as a pattern of closed loops in
the $z$ basis on the original lattice (red) or in the $x$ basis on the
dual lattice (green).  By blocking plaquettes of the
original lattice, we can obtain two representations: \textbf{(b)}~The
virtual indices double the loop degrees of freedom on the primal lattice,
and \textbf{(c)} in the resulting tensor, the virtual indices are the
difference of the adjacent physical indices.  \textbf{(d)}~The loops on
the dual lattice can be represented as differences of dual plaquette
colors, which form the virtual indices, and \textbf{(e)} in the resulting
tensor, the physical indices (in the dual basis) are the difference of
the adjacent virtual indices.}
\label{fig:tcode}
\end{figure}

We will now apply our framework to study the physics of
the Toric Code model with magnetic
fields,
\begin{equation}
\label{eq:H-TC-field}
H = H_{\mathrm{TC}} - h_x \sum_i \sx_i - h_z \sum_i \sz_i\ .
\end{equation}
Here, the degrees of freedom are two-level systems (qubits) sitting on the
edges of a square lattice, the sums run over all sites $i$, 
and 
\begin{equation}
H_{\mathrm{TC}}=-\sum_{p} (\sx)_p^{\otimes 4} - \sum_v (\sz)^{\otimes 4}_v
\end{equation}
is the Toric Code model~\cite{kitaev:toriccode},
where the sums run over all plaquettes $p$ and vertices $v$, respectively,
and $(\sx)^{\otimes 4}_p$ and $(\sz)^{\otimes 4}_v$ act on the four sites around
plaquette $p$ and vertex $v$, respectively, see Fig.~\ref{fig:tcode}a.

The Toric Code model exhibits $\mathbb Z_2$ topological order. Its ground
state minimizes all Hamiltonian terms individually and can either be seen
-- cf.\ Fig.~\ref{fig:tcode}a -- as an equal-weight superposition of all
loop configurations on the original lattice (solid lines) in the
$\sz$ basis $\{\ket0,\ket1\}$ (red loops), or of all loop
configurations on the dual lattice (dashed lines) in the $\sx$ basis
$\{\ket+,\ket-\}$ (green dashed loops). Its
ground state has an exact PEPS representation with $D=2$, and a $\mathbb
Z_2$ entanglement symmetry.  It can e.g.\ be derived in the
following two inequivalent ways, both relevant for later on: First, shown
in Fig.~\ref{fig:tcode}bc, 
by blocking the four
sites in every other plaquette to one tensor (gray square), 
``decorating'' the resulting
lattice as indicated (without adding physical degrees of freedom on the
additional edges), and defining the
decorated plaquette as one tensor -- that is, the virtual degrees of
freedom encode (in the $\{\ket0,\ket1\}$ basis) whether there is an 
outgoing loop at that point. 
Differently speaking, the tensor is constructed such that the 
virtual index is the difference (equivalently,
sum) modulo $2$ of the two adjacent physical indices.
Since only
closed loops appear, the $\mathbb Z_2$ entanglement symmetry precisely
corresponds to the fact 
that the number of loops leaving the tensor is even, i.e.\ there
are no broken loops.  We denote the generators of the symmetry group as
before by $Z$ (here, $Z=\sz$). 
In this representation, inserting a symmetry string corresponds to
assigning a $-1$ phase to all loop configurations which encircle the
endpoint of the string an odd number of times (a magnetic excitation, or
vison), while inserting a
non-trivial irrep such as
$X_\alpha=\left(\begin{smallmatrix}0&1\\1&0\end{smallmatrix}\right)$ or
$X_\alpha=\left(\begin{smallmatrix}0&1\\0&0\end{smallmatrix}\right)$
terminates a string and thus gives rise to broken strings (an electric
excitation).  Following the usual convention, we will denote the anyons by
$m\equiv (Z,1)$ and $e\equiv (\gid,-1)$, with $Z$ the non-trivial group element
of $G=\mathbb Z_2$.

Second, we can work in the dual loop picture (with loops in the
$\{\ket+,\ket-\}$ basis on the dual lattice), Fig.~\ref{fig:tcode}d, and assign ``color
variables'' to each plaquette such that loops are boundaries of colored
domains. If we choose the same blocking of four sites as before (gray
square), we obtain a tensor network representation where the virtual
indices carry the color label in the $\{\ket+,\ket-\}$ basis, and the
physical indices correspond to domain walls between colors, that is, the
tensor is constructed such that the
physical index is the difference modulo $2$ of the adjacent virtual
indices (all in the $\ket\pm$ basis), Fig.~\ref{fig:tcode}e. 
Here, the $\mathbb Z_2$ symmetry arises from the fact that
flipping all colors leaves the state invariant, and is thus again $Z\equiv
\sz$. In this dual basis, inserting an irrep $X_\alpha$ on a link assigns
a relative $-1$ phase to a colored plaquette (i.e.\ a plaquette 
enclosed by an odd number of loops
in the dual basis), while $Z$ strings flip
colors and thus break dual loops.

\subsection{Qualitative phase diagram}

What is the effect of a magnetic field on the Toric Code model?  If we
apply only a field $h_z>0$ in the $z$ direction ($h_x\equiv 0$), the field
commutes with the $(\sz)^{\otimes 4}_v$ term, and thus the ground state stays
within the closed loop space (on the original lattice). However, the field shifts the balance
between different loop configuration towards the vacuum configuration and
eventually induces a phase transition into a trivial phase.  This
disbalance between different loop configurations corresponds to a doping
with magnetic excitations, and thus, the phase transition is driven by
magnon condensation, while electric excitations become confined.  
(From now on, the terminology for excitations -- electric/magnetic etc.\ -- 
always refers to this basis, unless explicitly mentioned
otherwise.)
On the
other hand, a pure $x$-field $h_x>0$ has the same effect in the dual
loop basis but breaks loops in the $\sz$ basis, and thus 
induces a phase transition to a trivial phase through charge
condensation.    In fact, the whole model \eqref{eq:H-TC-field} has a
duality under exchanging $x$ and $z$ and at the same time going to the
dual lattice (which also exchanges electric and magnetic excitations), and
thus under $h_x\leftrightarrow h_z$.

\begin{figure}
\raggedright
\hspace*{2cm}\includegraphics[width=5cm]{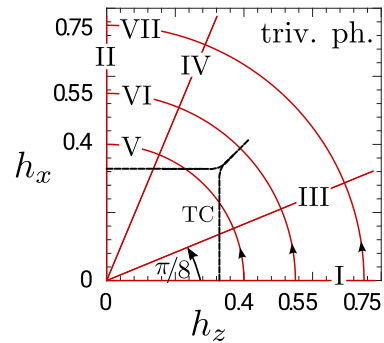}
\caption{Qualitative phase diagram of the Toric Code with $x$ and $z$ magnetic field,
Eq.~\eqref{eq:H-TC-field}. Phase boundaries are indicated in black, and
lines which we will study in detail later on in red, labeled by roman
numbers I--VII.  
There is a Toric Code phase (TC) at small field, which for large field
transitions into a
trivial phase either through flux condensation
($h_z>h_x$) or charge condensation ($h_x>h_z$). 
The model exhibits a duality under exchanging $h_x\leftrightarrow h_z$ 
and simultaneously electric and magnetic excitations.
Along the self-dual line $h_x=h_z$, there is a first-order line separating
the two different anyon condensation mechanisms through which the trivial
phase can be obtained, which ends at a sufficiently large field and is
replaced by a crossover regime.}
\label{fig:tc-phasediag}
\end{figure}

The phase diagram of the model is well
known~\cite{trebst:tcode-phase-transition,vidal:tcode-field-hx-hz,tupitsyn:tcode-multicritical,dusuel:tc-w-field,wu:tcode-field} and shown in
Fig.~\ref{fig:tc-phasediag} (where we mark lines which we are going to
study in detail with roman letters I--VII): There is a topological phase
at small field which transitions into a trivial phase through either flux
condensation (e.g.\ lines I and III) or charge condensation (lines II and
IV), as just discussed. Along the self-dual line $h_x=h_z$, there is a
first-order line which separates the charge condensed from the flux
condensed phase (crossed by line VI), which eventually disappears at large
enough field, at which point a crossover between the two different ways to
obtain the (ultimately identical) trivial phase through anyon condensation
appears (line VII). Along the two lines $h_x\equiv 0$ (line I) and its
dual $h_z\equiv 0$ (line II), it is well known that the ground state of
the model can be mapped to the ground state of the 2D transverse field
Ising model (we discuss the mapping in
Sec.~\ref{sec:TCfield:mapping-to-ising} in the context of our order
parameters).  Generally, the entire
transition line between the topological and trivial phase (except along
the diagonal) are believed to be in the 3D Ising universality class.

\subsection{Variational simulation\label{sec:tc:en-and-mag}}

For the iPEPS simulation, we work with the $2\times 2$ site unit cell
described above (Fig.~\ref{fig:tcode}bc) which contains one plaquette. 
We impose a virtual $\mathbb Z_2$ symmetry with generator
$Z=\openone_{D_+}\oplus (-\openone_{D_-})$, with $D=D_++D_-$ the bond
dimension. 
We optimize the variational energy by iteratively updating the tensor by
using Broyden-Fletcher-Goldfarb-Shanno (BFGS) algorithm
\cite{broyden1970convergence,fletcher1970new,goldfarb1970family,shanno1970conditioning}.
 After each update, we project the tensor back to the symmetric space.
To calculate the gradient of the objective function (i.e.\ the energy density)
 with respect to the tensor, we use the corner transfer matrix method
\cite{vanderstraeten2016gradient}. 
Furthermore, we observe that for the phase transitions between topological
and trivial phases, the BFGS algorithm always tends to converge faster and
find ground states with lower energies if it is initialized with the
tensor that belongs to the topological phase.  This observation suggests
an important feature of the optimization algorithm: As the algorithm
minimizes the energy by updating the local tensor at each step, it is
easier to remove than to build up long-range entanglement, and thus,
initializing with a state with more complex entanglement order is
advantageous.

\begin{figure}
\includegraphics[width=6cm]{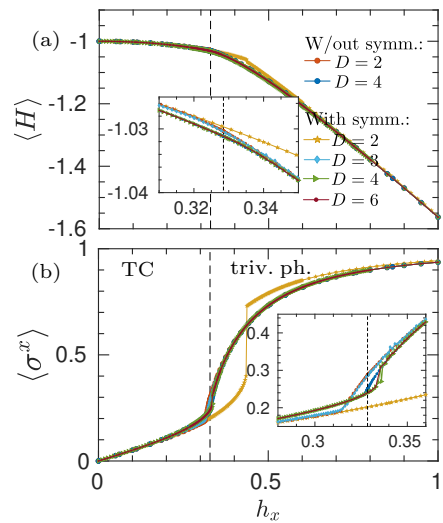}
\caption{Variational results for energy (a) and magnetization along the
field (b) for the Toric Code with an $x$ field.  We find
that for $D=4$, the results with symmetry are essentially fully
converged; on the
other hand, a simulation with $D=2$ with the entanglement symmetries
Eq.~\eqref{eq:G-injective} imposed yields a qualitatively wrong
first-order transition For comparison, we also show results obtained
without imposing symmetries.  See text for further details.}
\label{fig:tc-en-and-mag}
\end{figure}

Fig.~\ref{fig:tc-en-and-mag}a shows the variational energy obtained for an
$x$ field for $D=2,3,4,6$ (where $D=3=1+2=D_++D_-$, and otherwise
$D_+=D_-$), with the region around the critical point enlarged in the
inset.  We find that the optimal variational energy converges rather quickly with
$D$, with
energies for 
$D=4$ and $D=6$ already being indistinguishable.  
In addition, we observe that a symmetric splitting $D_+=D_-$ is generally
favorable.
For comparsion, we also
show energies obtained by optimizing PEPS tensors without any symmetry.
We find that $D=2$ without symmetries is comparable to $D=3$ with
symmetries (whereas $D=2$ with symmetries is considerably worse and in fact
gives a qualitatively wrong transition, as already observed in
Ref.~\cite{gu:TERG}), while $D=4$ with and without symmetry give essentially the
same energy. This demonstrates that imposing the symmetry does not
significantly restrict the variational space beyond halving the
number of parameters, and in particular, it does not necessitate to double the
bond dimension due to some non-trivial interplay of constraints.  Our
findings are also in line with previous observations that for the
transverse field Ising model (whose ground state is dual to ours),  the
energy is essentially fully converged for $D=3$~\cite{rader:peps-ising-scaling}.

In addition, Fig.~\ref{fig:tc-en-and-mag}b shows the magnetization along
the field. We see that for $D=2$ with symmetries, the phase transition is
off and first order.  For larger bond dimensions or without symmetries,
the point of the phase transitions is however rather close to the exact
value.  Notably, we see that the ansatz without symmetries undershoots the
critical point -- that is, it has a tendency towards the trivial phase --
while the ansatz with symmetries for $D\ge4$ slightly overshoots the
critical point -- that is, it has a tendency to stabilize topological
order.  Given the connection between entanglement symmetries and
topological order, this is indeed plausible.  An exception is the case of
$D=1+2$ with symmetries, which is closer to the $D=2$ case without
symmetries.  This indicates that the one-dimensional trivial irrep is
still too restrictive, and in this case, the ansatz possibly rather uses
the unrestricted degrees of freedom in the $2$-fold degenerate 
irrep space.

\subsection{Topological to trivial
transition: Order parameters}

\begin{figure}
\includegraphics[width=\columnwidth]{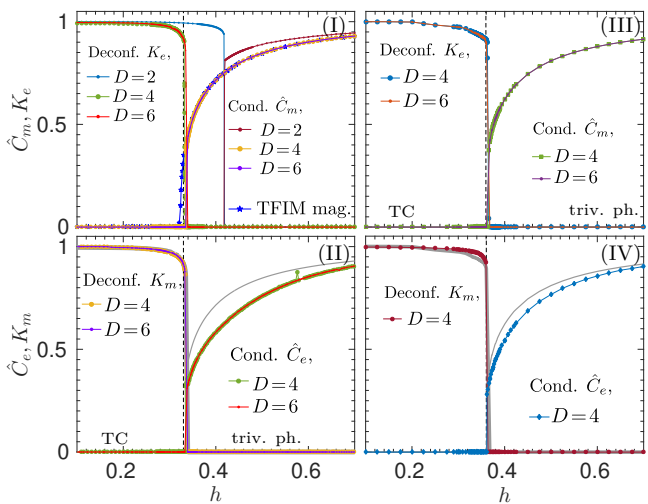}
\caption{Order parameters $\hat C_a$ for condensation and $K_a$ for
confinement across the four lines I-IV in Fig.~\ref{fig:tc-phasediag},
where along lines I and III, magnetic fluxes $a=m$ condense and charges $e$
confine, and vice versa for lines II and IV.  Even though I and II,
as well as III and IV, are dual to each other, the actual values of the order
parameters are different due to the gauge degree of freedom in the
construction of electric order parameters -- for comparison, the $D=6$
data from the first row is indicated in gray in the dual panels below. 
Yet, their critical exponents are the same, see
Figs.~\ref{fig:tc:exp-cond} and \ref{fig:tc:exp-conf}.
We
also observe that the magnetization of the transverse field Ising model
equals $\hat C_m$ along line I, as proven in 
Sec.~\ref{sec:TCfield:mapping-to-ising}.
\label{fig:tc:orderpars}}
\end{figure}

Let us first investigate the behavior of the order parameters as we
drive the system from the topological into the trivial phase by increasing
the field along a fixed direction.  Fig.~\ref{fig:tc:orderpars} shows the
order parameters for condensation and deconfinement for the four lines
I--IV.  Here, the first row reports the data for lines I and III, along which fluxes
condense, while the second row corresponds to lines II and IV, where
charges condense.  

Along all four lines, we observe a qualitatively similar behavior: As we
increase the field, the deconfinement fraction of the electric (I,III) or
magnetic (II,IV) charge decreases and drops to zero rather steeply at the
critical point, indicating their confinement.  
Past the critical point, the condensate fraction for the
condensed charge becomes non-zero, with an apparently much smaller slope. We
also see that the difference for the data with $D=4$ and $D=6$ is barely
visible, confirming what we found for the energy and magnetization in
Fig.~\ref{fig:tc-en-and-mag}. For line I (top left), we additionally show
the data for $D=2$: As already discussed in Section~\ref{sec:tc:en-and-mag},
it does not only give an incorrect critical point, but more importantly
also predicts a first- rather than second-order phase transition.

As discussed before, the lines I and II, as well as the lines III and IV
(each pair plotted 
in the same column),
 are self-dual to each other. On
the other hand, they clearly don't display the same value for the order
parameters, as can be seen from the lower panels (lines II and IV), where
we have indicated the $D=6$ data for their dual lines I and III as gray
lines. This is not surprising -- while the pairs of lines are dual to
each other, the way in which we extract the order parameters is not; in
particular, under the duality mapping the string-like order parameters,
which are gauge invariant, get mapped to the irrep-like order parameters,
which are not gauge invariant and require a gauge fixing procedure, and
vice versa. 

This non-uniqueness of the order parameters should not come as a
surprise, and is in fact in line with the discussion in
Sec.~\ref{sec:anyopar-quantitative-and-gauge},
where we discussed the ambiguities which arise in fixing an order parameter
when all we are allowed to use is the symmetry.  However, as we have
argued there, we expect that for well-designed order parameters (that is, a
well-designed gauge fixing procedure), we will observe the same universal
signatures, that is, the same critical exponents.

\subsection{Topological to trivial transition: Critical exponents}

Let us now study the scaling behavior of the order parameters in the
vicinity of the critical point.

\begin{figure}
\includegraphics[width=6.5cm]{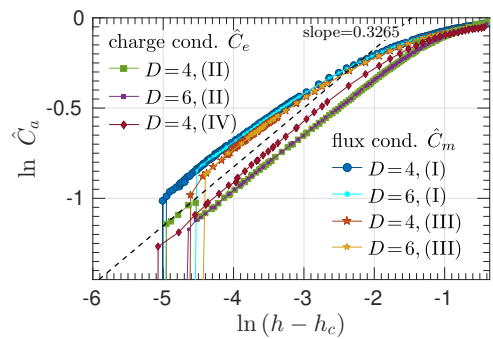}
\caption{Scaling of condensate fractions close to the critical point for
the lines I-IV.  The slope matches the critical exponent $\beta\approx
0.3265$ of the order parameter of the 3D Ising transition.}
\label{fig:tc:exp-cond}
\end{figure}

Fig.~\ref{fig:tc:exp-cond} shows the order parameter for anyon
condensation along the four lines I--IV (flux condensation for lines
I/III, charge condensation for lines II/IV).  We find that all lines show
the same critical scaling, which matches the known critical exponent
$\beta\approx0.3265$ of the magnetization in the (2+1)D Ising universality
class, consistent with the
fact that lines I and II map to the (2+1)D Ising model,  and confirming
the belief that the whole transition line is in the Ising universality
class.  Indeed, as we have observed in Fig.~\ref{fig:tc:orderpars},
the magnetic
condensate fraction along line I equals the magnetization in the (2+1)D
Ising model, a connection which will be made rigorous in
Sec.~\ref{sec:TCfield:mapping-to-ising}.

\begin{figure}
\includegraphics[width=6.5cm]{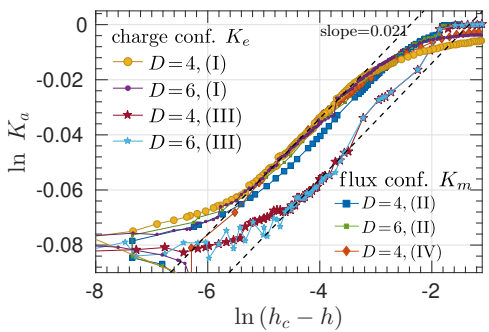}
\caption{Scaling of deconfinement fractions close to the critical point for
the lines I-IV.  The slopes along the different lines agree, yet give a
critical exponent $\beta^*\approx 0.021$, which is not among the known
critical exponents of the 3D Ising model. In the text, we discuss
interpretations of this exponents in terms of the 2D quantum Ising model,
the 3D classical Ising model, and the prefactor of the area law scaling of
the Wilson loop in a 3D Ising gauge theory.}
\label{fig:tc:exp-conf}
\end{figure}

Let us now turn towards the order parameter for deconfinement.
Fig.~\ref{fig:tc:exp-conf} shows the scaling behavior of the order
parameter for deconfinement along the same four transitions.  We again
find that the deconfinement fraction exhibits the same universal scaling
behavior along  all four lines. However, the critical exponent observed is
rather different, and much smaller, namely roughly $\beta^*\approx 0.021$.
However, the precise value should be taken with care, since (as always)
the fitting is rather susceptible to the value chosen for the critical
point, and the very small value of $\beta^*$ implies a rather large
relative error.

What is the nature of this new critical exponent $\beta^*\approx 0.021$,
which does not even in order of magnitude resemble any known critical exponent of the
(2+1)D Ising model?  In Sec.~\ref{sec:TCfield:mapping-to-ising}, we show
that the underlying order parameter maps to an order parameter obtained
from a ``twist defect line'' inserted into the ground state of the (2+1)D Ising
model which can be constructed based on its PEPS representation, and which
should serve as a disorder operator for the Ising model. This
suggests that our technique, developed with the characterization of
topological phase transitions in mind, can equally be used to construct novel
types of disorder parameters for \emph{conventional} phases. 
We
construct such a disorder parameter, and study it in detail 
for the (2+1)D Ising model,
in Section~\ref{sec:opar-conventional}, where we find that it indeed exhibits the same
novel critical exponent $\beta^*$. There, we  also discuss possible interpretations
of this critical exponent, as well as its utility in further characterizing the 
phase transition.

At the end of this section on critical exponents, let us stress
that the fact that our order parameters give the same universal
behavior, even though the dual order parameters for the charge and
flux condensation transition are constructed in entirely different ways
(in particular, charges require gauge fixing, while fluxes don't) gives an
a posteriori confirmation of our approach to extracting order parameters
and universal behavior.

\subsection{Topological to trivial transition: Anyon masses}

\begin{figure}[t]
\includegraphics[width=6.5cm]{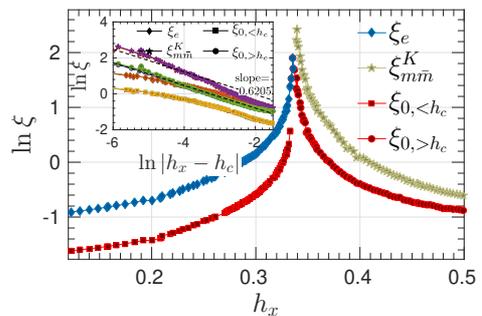}
\caption{Scaling of different correlation lengths along the line~II:
Inverse mass gap $\xi_e$ for charges (in the topological phase),
confinement length $\xi^K_{m\bar m}$ for fluxes (in the trivial phase),
and trivial correlation length $\xi_0$. The scaling analysis (inset) shows
that they all exhibit the same critical exponent, which matches that of
the 3D Ising transition.}
\label{fig:tc-massgap}
\end{figure}

As discussed, we can also extract length scales from our simulations.
Specifically, we can on the one hand extract 
correlation lengths $\xi_a$ for anyon-anyon correlations, 
or, equivalently, anyon masses $m_a=1/\xi_a$, for free anyons; a divergence of $\xi_a$
(i.e., a closing mass gap) witnesses a condensation of anyon $a$. On the
other hand, we can extract a confinement length scale $\xi^K_{a\bar a}$
for confined anyons, which diverges as the anyons become deconfined.

Fig.~\ref{fig:tc-massgap} shows these lengths along the line II, where
charges condense.  Specifically, we see that the inverse anyon mass of the
electric charge, $\xi_e$, diverges at the phase transition, while in the
trivial phase, the magnetic fluxes become confined, witnessed by a finite
confinement length $\xi^K_{m\bar m}$. In addition, we also show the
inverse mass gap for topologically trivial excitations, 
which diverges at the critical
point as well, but is smaller than the other length (typically, one would
assume that trivial excitation with the smallest mass gap is constructed from
a pair of topological excitations, and thus should have roughly twice
their mass, neglecting interactions).  

The analysis of the critical scaling of the different lengths reveals that
they all display the same scaling behavior, consistent with the critical
exponent $\nu$ of the correlation length in the (2+1)D Ising model.

\subsection{Rotating the direction of the magnetic field: First-order line and crossover}

\begin{figure}[t]
\includegraphics[width=\columnwidth]{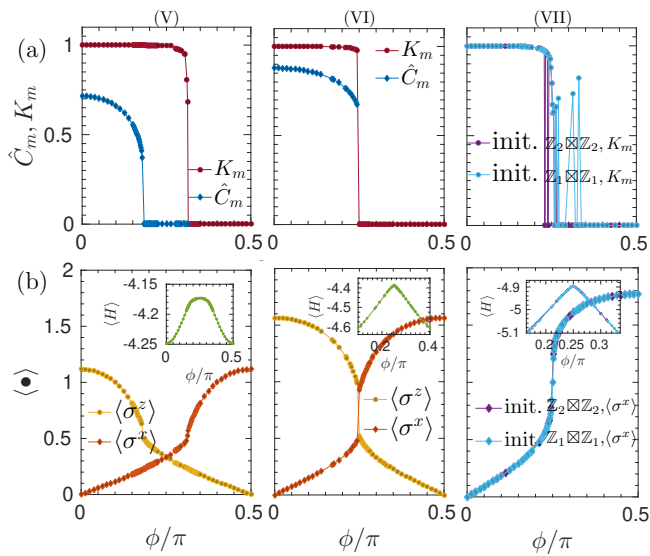}
\caption{Behavior for rotating field, lines V--VII, which move between
two different condensation mechanisms of realizing the trivial phase. Each
column corresponds to one of the lines V--VII.
Top row: Condensate and deconfinement fraction of magnetic fluxes.  Bottom
row: Magnetization $\langle\sigma_x\rangle$ and $\langle\sigma_z\rangle$,
and energy $\langle H\rangle$ (inset). Line V has two second-order phase
transitions with a topological phase in between, line VI a first-order
phase transition between the two inequivalent trivial phases, and line VII
a crossover. For line VII, we rather show the deconfinement fraction, the
$x$ magnetization, and the energy for two different choices of initial
conditions (following the notation of
Ref.~\cite{duivenvoorden:anyon-condensation}, $\mathbb
Z_1\boxtimes\mathbb Z_1$ denotes the flux confined phase and $\mathbb
Z_2\boxtimes \mathbb Z_2$ the flux condensed phase): We find that while
the physical properties converge independent of the initial configuration,
the interpretation in terms of a charge or flux condensate becomes
unstable around $\phi\approx\pi/4$, indicating a crossover regime where
the interpretation of the physical phase in terms of the virtual
symmetries breaks down.}
\label{fig:tc-V-VI-VII}
\end{figure}

Finally, let us study what happens when we rotate the magnetic field in
the $x$-$z$-plane while keeping its strength constant, i.e., moving
radially in the phase diagram Fig.~\ref{fig:tc-phasediag} along the three
lines V, VI, and VII.   The resulting data is shown in
Fig.~\ref{fig:tc-V-VI-VII}.  Here, the panels in the first line show
the condensation and deconfinement fractions for the magnetic particles,
while the panels in the second line display the
behavior of the $x$ and $z$ magnetization as a function of the angle
$\phi$, with the energy shown in the inset. 
The three columns
correspond to the three radial lines V, VI, and VII.

For the line V, we observe two second-order topological phase
transitions, first from the trivial to the topological phase through
decondensation of the magnetic flux, and subsequently from the topological
to the trivial phase through flux confinement. Both $\hat C_m$ and $K_m$
show a clear second-order behavior.  Similarly, the two magnetizations
$\langle \sx\rangle$ and $\langle \sz\rangle$ each show a kink,
yet again indicative of underlying second-order transitions.  On the other
hand, the energy does not exhibit clear signs of the phase transitions,
which will only show up in its derivatives.

For the line VI, the condensation and the confinement of the magnetic
flux coincide at $\phi=\pi/4$: The system undergoes a transition from a flux condensed to
a charge condensed (flux confined) phase, without going through an
intermediate topological phase. In addition, the order parameters $\hat
C_m$ and $K_m$ show a clear jump, indicative of a first-order transition.
Similarly, $\langle \sx\rangle$ and $\langle \sz\rangle$ both
exhibit a discontinuity at $\phi=\pi/4$, and the energy
shows a kink (and thus a discontinuous derivative).

Finally, along the line VII, the order parameter plot now shows two
curves for the deconfinement fraction $K_m$, obtained by starting the
optimization from two different initial states, either in the charge or in
the flux condensed phase. 
 We see that around $\phi\approx \pi/4$, the value
of the deconfinement fraction becomes unstable and depends on the choice
of the initial phase. This is not all too surprising, since the line VII
realizes a crossover between the two different mechanisms of realizing the
trivial phase, and in the crossover regime, the interpretation of the trivial
phase as an either charge or flux confined phase should become ambiguous;
the observed dependence of the deconfinement fraction $K_m$ on the initial
phase can thus be taken as a fingerprint of this crossover.  On the
other hand, the lower panel shows that the \emph{physical} state
obtained in the optimization is \emph{stable} independent of the choice of
the initial condition: Both the value of $\langle \sx\rangle$ and the
energy are independent of the choice of the initial tensor.  The observed
instability is thus purely a signature of the ambiguous \emph{interpretation}
of the trivial phase in the crossover regime when thought of as a
condensed version of the topological model -- that is, the way the state
is realized on the entanglement level -- rather than an instability of
the variational method as such.

\subsection{Mapping to the (2+1)D Ising 
model\label{sec:TCfield:mapping-to-ising}}

It is well known that there is an analytical mapping of the ground state
of the Toric Code with only an $x$ or a $z$ field to the (2+1)D Ising
model (i.e., the 2D transverse field Ising
model)~\cite{trebst:tcode-phase-transition}. In the following, we
will use this mapping to interpret our order parameters for
condensation and confinement in terms of conventional and generalized
order parameters for the (2+1)D Ising model.

\begin{figure}
\includegraphics{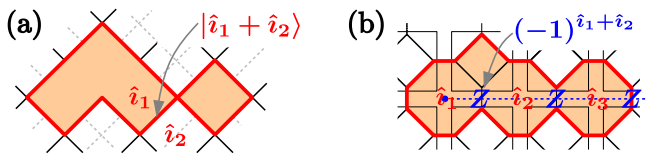}
\caption{Mapping of the Toric Code with field to the Ising model; the
mapping works on the space of closed loops.
\textbf{(a)}~The Ising variables are obtained by assigning color labels
$\hat \imath$ to the plaquettes, where loops are the domain walls between
different colors. \textbf{(b)} $Z$ operators measure the difference between
two adjacent colors, $(-1)^{\hat\imath_1+\hat\imath_2}$. A magnetic flux
($Z$ string) thus corresponds to a $z$ correlator
$(-1)^{\hat\imath_1+\hat\imath_\ell}$ between the Ising variables at its
ends.}
\label{fig:opar-dual-x}
\end{figure}

To this end, we start by briefly reviewing the mapping.  To start
with, consider the Toric Code with a $z$ field,
\begin{equation}
\label{eq:H-TC-z-field}
H = 
-\sum_{p} (\sx)_p^{\otimes 4} - \sum_v (\sz)^{\otimes 4}_v
- h_z \sum_i \sz_i\ .
\end{equation}
Since the field $\sz_i$ commutes with the vertex stabilizers
$(\sz)^{\otimes 4}_v$, for any $h_z$ the ground state is spanned by closed
loop configurations in the $\{\ket0,\ket1\}$ basis on the original lattice.  We can thus
work in a dual description of the loop basis, similar to
Fig.~\ref{fig:tcode}d, but now on the original lattice, where we color
plaquettes $p$ with two colors (white=$\ket0$, red=$\ket1$), 
and interpret loops as domain walls of color domains, see
Fig.~\ref{fig:opar-dual-x}a.
We will label plaquette variables by $\pket{\hat\imath_p}$ and also mark
Hamiltonian terms (Paulis) acting on them by a hat.

Let us now see how the Hamiltonian \eqref{eq:H-TC-z-field} acts in the
dual basis.
The Hamiltonian term $(\sz)_v^{\otimes 4}$ is then trivially satisfied.
$(\sx)_p^{\otimes 4}$ flips the loop around $p$, and thus corresponds to
flipping the plaquette color $\ket{\hat\imath_p}$, i.e., it acts as
$\hsx_p$. On the other hand, the magnetic field $\sz_i$ assigns a sign
$-1$ to a loop on that edge; as loops are domain walls of
plaquette colors, this corresponds to 
$(-1)^{\hat\imath_p+\hat\imath_p'}$ and thus $\hsz_p\hsz_{p'}$. In this
basis, $H$ (restricted to the loop space, i.e., the ground space of
$\sum_v(\sz)_v^{\otimes 4}$) thus becomes 
\begin{equation}
\hat H = -\sum_p\hsx_p - h_z\sum_{\langle
p,p'\rangle}\hsz_p\hsz_{p'}\ .
\end{equation}
Note again that this is primarily a mapping between the ground states of
the models and in particular does not cover excitations beyond the closed
loop space.

Let us now see what happens to the anyonic order parameters under this
mapping.  We will focus our initial discussion on the order parameters
constructed from $Z$ strings, since  these are gauge
invariant and thus yield a unique quantity on the dual Ising model.
However, the mapping can also be applied to irrep-like order parameters
$X_\alpha$, and we will give a brief account of those at the end of the
discussion.

First, let us consider the case of a $z$-field as just discussed.
 In that case, the natural tensor network representation -- that is,
the one which is constructed from the loop constraint in the $z$ basis -- 
is the one in Fig.~\ref{fig:tcode}c. The key property lies in the fact
that the irreps on the virtual legs carry the loop constraint (that is,
the irrep label of the virtual index equals the sum of the adjacent
physical legs in the loop basis).  As it turns out,
this property is preserved by the variationally optimal wavefunction also
at finite field, and thus, anyonic order parameters constructed on the
entanglement level still have a natural interpretation in terms of the
loop picture, and thus of the dual Ising variables. We have verified 
numerically that this holds to high accuracy, 
but it is also plausible analytically: On the
one hand, the ground state is constrained to the closed loop space, and on
the other hand, the tensor is constrained to the $\mathbb Z_2$-symmetric
space, and thus, identifying these two constraints should give the
maximum number of unconstrained variables to optimize the wavefunction.

Now consider the order parameter for condensation, that is, a
semi-infinite (or very long finite) $Z$ string, see
Fig.~\ref{fig:opar-dual-x}b.
This $Z$ assigns a $-1$ sign to every edge with a loop,
and thus for every edge, its effect equals to $(-1)^{\hat\imath_1+\hat\imath_{2}}$
for the two adjacent plaquettes $1$ and $2$, as indicated in
Fig.~\ref{fig:opar-dual-x}b. Thus, for a long string of $Z$'s, the overall action
equals 
$(-1)^{\hat\imath_1+\hat\imath_2}
(-1)^{\hat\imath_2+\hat\imath_3}\cdots
(-1)^{\hat\imath_{\ell-1}+\hat\imath_\ell}=
(-1)^{\hat\imath_1+\hat\imath_\ell}$, 
and thus the two-point correlator
$\hsz_1\hsz_\ell$ of the Ising model variables.  As the condensation order
parameter evaluates the overlap of this state with the ground state, it
measures $\langle\hsz_1\hsz_\ell\rangle$:  We thus find that
the order parameter for flux condensation under a $z$ field maps
precisely to the magnetization in the 2D transverse field Ising model --
as we have already observed numerically in Fig.~\ref{fig:tc:orderpars}.

Let us now turn to the case of the $x$ field. Here, 
the ``good'' basis is the one spanned by $x$ basis
loops on the dual lattice, and thus, we naturally arrive at the tensor
network representation Fig.~\ref{fig:tcode}e. Its defining feature -- which
we have again checked numerically to also hold away from the Toric Code point --
is that the loops, that is, the physical degrees of freedom in the
$\{\ket+,\ket-\}$ basis, are obtained as the difference of the ``color''
label of the virtual legs.  However, different from before, the color
label is not uniquely defined: ``Color'' corresponds to a decomposition 
of the bond space as
$\mathbb C^D \simeq \mathcal S_\mathrm{white}\bigoplus \mathcal
S_{\mathrm{green}}$,  such that $Z$
acts by swapping the two color spaces, $Z\,
\mathcal S_\mathrm{white}=\mathcal S_{\mathrm{green}}$. Indeed, by applying any 
matrix $\Lambda$ which commutes with $Z$, we can obtain another such decomposition
(even with a non-orthogonal direct sum). This ambiguity in the choice of
the color basis -- which becomes
precisely the Ising basis after the duality
mapping -- is a reflection of the fact that in our approach, the only basis
fixing comes from the symmetry action, leaving room for ambiguity, as
discussed in Section~\ref{sec:topo-opar}.  However, let us point out that numerically
we observe that the ``physical index equals difference of colors''
constraint is very well preserved for the ``virtual $x$ basis'', that is, for the
``color projections'' $\left(\begin{smallmatrix}\openone&\pm\openone\\
\pm\openone&\openone\end{smallmatrix}\right)$, likely due to the choice of
initial conditions (the Toric Code tensor) in the optimization.

\begin{figure}[t]
\includegraphics[width=7cm]{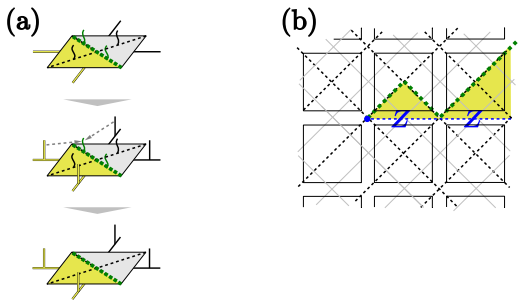}
\caption{\textbf{(a)} Mapping from the PEPS tensor for the Toric Code in a
field in the dual representation of Fig.~\ref{fig:tcode}e (top) to the
Ising model. First, the
virtual degrees of freedom (color variables) are copied to physical
degrees of freedom, which will become the Ising degrees of freedom (middle;
three meeting lines correspond to a delta-tensor in the Ising = $\ket\pm$ basis).
Then, the physical degrees of freedom of the Toric Code are disentangled, e.g.\ by using
controlled-not operations in the $\ket\pm$
basis, controlled by the Ising degrees of freedom (as indicated by the
arrows in the middle panel). The disentangled degrees of freedom can then
be discarded, resulting in a tensor network for the Ising model (bottom).
This can be seen as the reverse of a gauging transformation.
\textbf{(b)} Effect of a string of $Z$ operators in the dual
representation: $Z$ operators flip the color label of a plaquette, giving
rise to a domain wall in the coloring and thus broken dual loops (green
line) at the endpoint of the $Z$ string.  
\label{fig:opar-duality-conf}}
\end{figure}

Since in this PEPS representation, the Ising degree of freedom in the
duality mapping is nothing but the color degree of freedom of the
plaquettes, the mapping from the Toric Code to the Ising model can be made
very explicit on the level of the PEPS: We need to 
duplicate the color degree of freedom as a physical degree of
freedom, and subsequently remove the original physical degrees of freedom of the
Toric Code, similar to an ungauging procedure, see
Fig.~\ref{fig:opar-duality-conf}a.  The latter can be done,
for instance, through a controlled unitary (in the dual basis) controlled by the Ising (color)
degrees of freedom, since we know that the physical degrees of freedom are
just their differences.  Note that for this construction to work, we must
know the correct color basis (see above), which however is a property
which can be extracted from the tensor (and is only needed in case we want
to carry out the mapping explicitly).

For an $x$ field, at the phase transition fluxes become confined.  What
does the order parameter for flux deconfinement -- the normalization
of the PEPS with a semi-infinite (or very long) string of $Z$'s placed
along a cut -- get mapped to in the Ising model?  The effect of a  $Z$ is to
flip the color label.  A semi-infinite string of $Z$'s thus flips the
color labels along a semi-infinite cut on the lattice. Since the loop
variables are the difference of the color variables, and the ``closed
loop'' constraint is implicitly guaranteed by the fact that we arrive at
the same color when following a closed curve on the original lattice (recall that
the loops live on the dual lattice, and thus the colors on the vertices
of the original lattice), flipping the color variable within
the plaquette gives rise to a broken ``closed loop'' constraint for any
circle around the endpoint of the $Z$ string -- that is, the endpoint of
the $Z$ string is the endpoint of a broken loop, see
Fig.~\ref{fig:opar-duality-conf}b.  Indeed, this is 
precisely what a magnetic flux corresponds to in the dual basis: a
broken string.

However, how can this be mapped to the Ising model? The fact is that it
cannot, at least not in a direct way which gives rise to an observable for the
Ising model: The mapping to the Ising model
precisely relies on the fact that we are in the closed loop space, which
is no longer the case in the current basis after introducing a flux. However,
we can still give an interpretation of this object in 
terms of the Ising model,
if we describe the ground state of the Ising model in terms of PEPS:
After the duality mapping described above, we obtain
nothing but a variational PEPS description of the ground state of the
Ising model (which becomes exact as the bond dimension grows), constructed
from tensors with a $\mathbb Z_2$ symmetry
\begin{equation}
\raisebox{-0.7cm}{\includegraphics{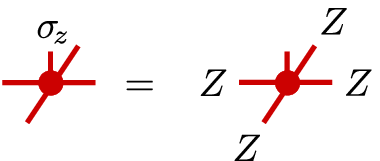}}
\end{equation}
(e.g.\ by blocking the ``ungauged'' tensor at the bottom of
Fig.~\ref{fig:opar-duality-conf}a with the left and bottom physical index).
The order parameter then corresponds to inserting a semi-infinite string
of $Z$'s along a cut -- a ``twist defect'' -- and computing the normalization of the modified
tensor network (relative to the original one). 
It can be easily seen that
this is zero in the ordered phase: In that case, the virtual indices
carry the information about the symmetry broken sector, that is, they
are all supported predominantly in the same sector, which is flipped by
the action of the $Z$ string. Glueing the network with a $Z$ string thus
leads to a decrease in normalization which goes down exponentially with
the length of the string, as configurations which are approximately in
different sectors (with overlap $<1$) are being glued together.  Conversely, in the disordered
phase we generally expect a non-zero norm, since
sufficiently far away from the cut, the spins will be disordered and thus
not have a preferred alignment relative to each other along the cut.  The
only contribution comes from the endpoint of the string (since the spins
are still aligned up to a scale on the order of the correlation length).
 Thus, we expect a non-zero value in the disordered
phase and a zero value in the ordered phase (a \emph{disorder parameter}),
and thus a non-trivial behavior as the phase transition is approached.

It is notable that this way, we can define a (dis-)order parameter 
 for the Ising model based on its ground state,
even though there is no direct way of measuring it from the ground state
itself: Rather, one first has to find a $\mathbb Z_2$-symmetric PEPS
representation of the ground state and construct the order parameter 
through the effect of 
twisting the PEPS on the entanglement degrees of freedom. 
In some sense, it is the \emph{combination} of the correlation structure of the ground state
and the locality notion imposed 
by the tensor network description on the quantum correlations
which makes this possible.
This is the reason why the deconfinement order parameter allows us to
transgress the mapping to the Ising model, and thus probe properties of
the system which are inaccessible when directly probing the system.
Let us note that of course, the twisted state is no longer a ground state
of the Ising model, and has a large energy around the twist, which however
yet again reinforces the point that this type of order parameter is
defined through a deformation of the tensor network description of the
ground state, and not as a directly observable  property of the ground
state as such.

This discussion suggests that the same ideas as used in the construction
of topological order parameters can also be applied to directly construct disorder
operators for phases with conventional order, such as the (2+1)D Ising
model; this is studied in detail in Section~\ref{sec:opar-conventional}.

Finally, an analogous mapping can be carried out for electric charges.
In the case of the $z$ field, a charge breaks a loop, and correspondingly,
the duality mapping to the Ising model via plaquette colors breaks down.
This can be remedied by introducing a twist along a line emerging from the
charge across which the color, that is, the Ising variable, is flipped;
this gives rise to precisely the same order parameter constructed from
inserting a twist defect in the PEPS representation of the Ising ground
state as
discussed above.  In the case of the $x$ field, on the other hand, the
charge operator maps directly to the magnetization operator of the Ising
(color) variable, given that it is constructed in the right way relative
to the good color basis $\mathcal S_{\mathrm{white}}\oplus\mathcal
S_\mathrm{green}$.

\section{Entanglement order parameters for conventional phase
transitions\label{sec:opar-conventional}}

The fact that, under the duality mapping between the Toric Code 
the transverse field Ising model, the deconfinement order parameter 
gets mapped to a twist defect in the Ising model, which should
serve as a disorder parameter, suggests the surprising possibility that
tensor networks and the direct access to entanglement which they provide
can be used to also construct disorder parameters for \emph{conventional}
phase transitions in entangled quantum matter; and the unexpected critical
exponent observed for the deconfinement fraction suggests that this might
equally give access to new critical exponents for those conventional
transitions.

In the following, we describe a general such framework for the
construction of disorder operators for conventional phase transitions,
based on the direct access to entanglement provided by tensor networks. We
then present numerical results obtained for the (2+1)D transverse field
Ising model which confirm that this new kind of disorder operator indeed
displays a new critical exponent which is not found otherwise in the 3D
Ising theory. We discuss its relation to other disorder operators,
and we conclude by explaining how both these disorder parameters and order
parameters for conventional phase transitions, as well as the order
parameters for topological phase
transitions which we constructed, can be understood on a unified footing as entanglement order
parameters, that is,  order parameters which are constructed to identify
(dis)ordering relative to \emph{all} symmetries of the system -- 
physical as well as entanglement symmetries -- on a unified footing.

\subsection{Construction}

Consider a Hamiltonian with a global symmetry,
$[H(\lambda),U_g^{\otimes N}]=0$, where $U_g$ is a (faithful)
representation of some symmetry group $G$. We will focus on discrete
symmetry groups $G$ in the following, though most of our arguments (with some
caveats) apply to continuous groups as well. As $\lambda$ 
changes, the Hamiltonian undergoes a phase transition from a
symmetry-broken (ordered) to a symmetric (disordered) phase. As an
example, one could e.g.\ think of the transverse field Ising model or a
$\mathbb Z_N$ Potts or compass model with magnetic fields.  

Let us now approximate the ground state (using a variational 
method) with a tensor network state $\ket\Psi$ of some bond dimension $D$, where the
symmetry has been encoded in the local tensor $A$,
\begin{equation}
\raisebox{-.65cm}{\includegraphics[scale=.9]{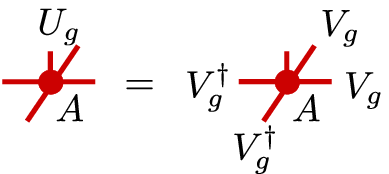}}\quad .
\end{equation}
From the tensor network $\ket\Psi$, we can of course compute 
conventional order parameters, i.e., measure the expectation value of a
local physical operator $S$ which does not commute with $U_g$, such
as an operator $S_\alpha$ transforming as a non-trivial
one-dimensional irrep $\alpha$ of $G$, $U_gS_\alpha =
\alpha(g)\,S_\alpha U_g$.

However, from the tensor network for $\ket\Psi$ we can also define a
disorder parameter as follows: First, define a state 
$\ket{\Psi_g(\ell)}$ by inserting a ``twist'' with a string of $V_g$'s of
length $\ell$, 
\begin{equation}
\ket{\Psi^\mathrm{tw}_g(\ell)} = 
\raisebox{-1.15cm}{\includegraphics[scale=.9]{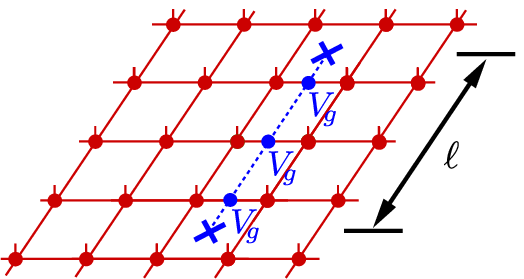}} .
\end{equation}
Then, define
\begin{equation}
N_g(\ell) =
\frac{\langle\Psi^\mathrm{tw}_g(\ell)\ket{\Psi^\mathrm{tw}_g(\ell)}}{\langle\Psi(\ell)\ket{\Psi(\ell)}}\ .
\end{equation}
What behavior do we expect $N_g(\ell)$  to display in the ordered and
disordered phase, respectively? To get a qualitative understanding, let us consider the case of
$G=\mathbb{Z}_N\cong\{0,1,\dots,N-1\}$ with a representation $U_g=\sum 
\ket{g+j}\bra{j}$. In the limiting case of a Potts model with zero field,
the ground space is spanned by states of the form $\sum w_j
\ket{j}^{\otimes N}$. The corresponding tensor $A$ (with $V_g=U_g$) is
then a tensor with all indices the same, $A=\sum_j
w'_j\ket{j}\bra{j,j,j,j}$, with some suitable weight $w_j'$. This forces
all physical \emph{and} virtual indices in any connected component  of the tensor network 
 to have the same value $j$. However, when placing a twist $V_g$
($g\ne 0$) along a
cut, $j$ is changed to $g+j$, which is orthogonal to the virtual index on
the other side of $V_g$, and thus, $N_g(\ell)=0$. On the other hand, in
the limit of infinite field, the ground state is of the form
$\ket{+}^{\otimes N}$, $\ket+=\sum\ket g$, which can be represented by a
tensor $A=\ket{+}\bra{t,t,t,t}$, where $\ket{t}$ is a state in the trivial
irrep of $V_g$. Thus, $V_g$ acts trivially on the tensor, and
$N_g(\ell)=1$.

\begin{figure}[t]
\includegraphics[width=220pt]{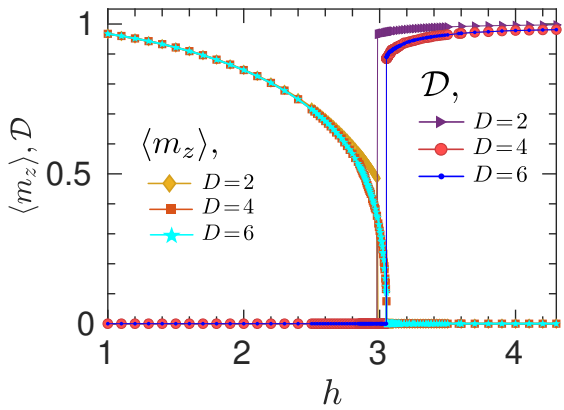}
\caption{Order parameter  $\langle m_z\rangle$ 
($z$ magnetization)
and
disorder parameter $\mathcal D$ (response of normalization to inserting a semi-infinite
``symmetry twist'') for the (2+1)D transverse field Ising model; see text
for details.
\label{fig:ising-opars}}
\end{figure}

As we interpolate between the two phases, we expect that $N_g(\ell)$
interpolates between these two behaviors. In the ordered phase, we expect
that the physical degrees of freedom, and thus also the entanglement
degrees of freedom, are aligned up to short-ranged fluctuations on the
order of the underlying length scale, and thus, we expect $N_g(\ell)\sim
e^{-\ell/\xi}$, where $\xi$ has a critical exponent $\nu^*=\nu$. On the
other hand, in the disordered phase, we expect that the spins are not
correlated beyond the correlation length, and thus, $N_g(\ell)\to \mathcal D^2>0$.
Thus, $\mathcal D$ serves as an order parameter for the disordered phase, that is,
it is a disorder parameter (also called disorder operator). Note that
$\mathcal D$ can be 
considered as the normalization of the ground state tensor network with a
semi-infinite twist, given suitable boundary conditions.

\subsection{Numerical results}

Let us now study how this disorder parameter behaves for a model with
a symmetry breaking phase transition. Specifically, we consider the 2D
transverse field Ising model
\begin{equation}
H_\mathrm{Ising} = -\sum_{\langle i,j\rangle} \sigma_i^z\sigma_j^z -h\sum_i
\sigma_i^x\ .
\end{equation}
It possesses a $\mathbb Z_2\equiv\{0,1\}$ symmetry
$[H_\mathrm{Ising},U_g^{\otimes N}]=0$ with $U_1=\sigma^x$.
We variationally optimize a tensor network ansatz for the ground state of
$H_\mathrm{Ising}$ with a $\mathbb Z_2$ symmetry encoded,
\begin{equation}
\raisebox{-.65cm}{\includegraphics[scale=.9]{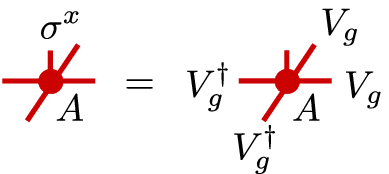}}\quad
,
\end{equation}
where $V_1=\sigma_x\otimes \openone_{D/2}$, using the same numerical methods as
described in Section~\ref{sec:tc:en-and-mag}.

\begin{figure}
\hspace*{-30pt}\includegraphics[width=280pt]{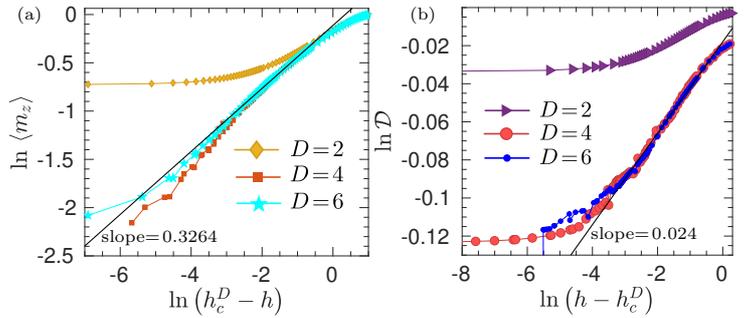}
\caption{Critical scaling for  order parameter and  disorder
parameter at the phase transition. As for the Toric Code, we find that
convergence is reached starting from bond dimension $D=4$, while $D=2$ is overly
restricted due to the symmetries. \textbf{(a)} For the order parameter
$\langle m_z\rangle$, we
recover the well-known scaling of the 3D Ising unversality class.
\textbf{(b)} For the disorder parameter $\mathcal D$, we observe a new critical
exponent $\beta^*\approx 0.024$, which matches the critical exponent
observed for the deconfinement fraction in the Toric Code model.
\label{fig:ising-exponents}}
\end{figure}

From the optimized tensor, we have determined both the order parameter
(the magnetization) and the disorder paramater (the normalization of the
twist defect). Fig.~\ref{fig:ising-opars} shows the results: As expected, we find that in the
ordered phase, the order parameter is non-zero and the disorder parameter
is zero, and vice versa in the disordered phase. As for the previously
considered Toric Code model with magnetic fields, we observe that the
disorder parameter vanishes much more steeply as the phase transition is
approached. 
Fig.~\ref{fig:ising-exponents} shows the scaling of the order and disorder parameter as the phase
transition is approached. For the order parameter, our data is in
agreement with the well-known value 
$\beta\approx 0.3265$, while for the disorder parameter, we find that it vanishes with a critical
exponent of about $\beta^* \approx 0.024$, fully compatible with what
we observed for the deconfinement fraction in the Toric Code model.

\subsection{Relation to other disorder parameters}

\begin{figure}
\includegraphics{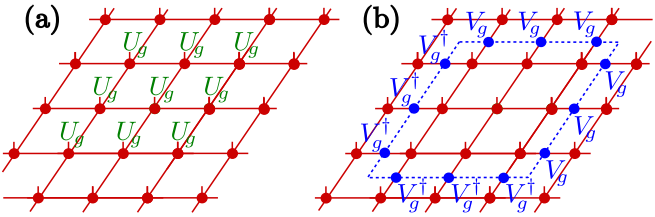}
\caption{Disorder parameter constructed from a physical symmetry action on
a membrane (panel a) and its reformulation in terms of an entanglement
symmetry (panel b), see text for details.
\label{fig:membrane}}
\end{figure}

Recently, another way of constructing disorder parameters has been
proposed, namely to act with a membrane of physical symmetry operators
$U_g^{\otimes \mathcal R}$ on a region $\mathcal R$ of the ground state $\ket\Psi$,
as shown in Fig.~\ref{fig:membrane}a, and to compute the overlap with the ground
state, i.e., $\Theta:=\bra\Psi U_g^{\otimes M}\ket\Psi$. In the
ordered phase, this will lead to a state which is approximately orthogonal
to $\ket\Psi$ locally in all of $\mathcal R$, and thus, one expects 
a volume law scaling $-\log\Theta
\sim c\,|\mathcal R|$ (here, $|\mathcal R|$ is the volume of $\mathcal R$).
On the other hand,  in the 
disordered phase, acting with $U_g^{\otimes\mathcal R}$ will only have an
effect on the the boundary of $\mathcal R$ but not on its (disordered) bulk,
and thus we expect a boundary law scaling $-\log\Theta \sim
d\,|\partial\mathcal R|$ (with $|\partial\mathcal R|$ the length of the
boundary of $\mathcal R$).
Specifically, as the phase transition into the ordered phase is
approached, $d$ must diverge in order to transition to a volume law
scaling, and thus, $d^{-1}$ can serve as an order parameter.
Indeed,
this is what was observed in Ref.~\cite{zhao:ising-disorder-operator} for the transverse field Ising
model, and in particular, it was found that $d$ scales as the correlation
length $\xi$ and thus, $d^{-1}$ vanishes with the same critical exponent
$\nu\approx 0.6205$.

It is remarkable that these two different ways to define disorder
parameters result in such different scaling behaviors. Are these two order
parameters entirely unrelated? To start with, note that if a tensor
network with symmetric tensors is used, it is immediate that a physical
membrane $U_g^{\otimes \mathcal R}$ is equivalent to a virtual loop
operator $V_g^{\otimes \partial \mathcal R}$ as shown in
Fig.~\ref{fig:membrane}b.  
Thus, $e^d$ can be interpreted as the \emph{overlap per unit length} of the
tensor network with and without an infinite (not: semi-infinite!) twist line. We thus see that
both order parameters share quite some similarity: They are both obtained
by measuring the effect of inserting
twist defects in the tensor network. However,  they are also 
rather distinct in other ways: While our disorder parameter is obtained
from an \emph{open-ended} string, the other is obtained from a
\emph{closed} or infinite string. Moreover, in one case the order
parameter is a normalization, while in the other, it is an overlap.
And finally, one of them has a
length-dependent contribution and needs to be taken per unit length,
while the other one doesn't display such a behavior (that is, it is
length-independent), an distinction clearly confirmed
by the numerics.

It is nevertheless tempting to think that the two order parameters should,
in fact, behave the same way, by using a Wick rotation argument. The
pronounced difference observed between these two order parameters in the
numerical simulations, however, makes it clear that this is not the case.
Let us nevertheless briefly discuss the Wick rotation argument, and also
why one shouldn't expect such a behavior.  Specifically, we can think of
the ground state of a Hamiltonian as being
obtained from imaginary time evolution $e^{-\beta H}$ of an arbitrary
initial state; through trotterization, one arrives at a 3D tensor network for
the ground state as well as expectation values, which can at the same time
be seen as a way to construct the 2D tensor network by blocking
columns.  Specifically, for the Ising model, one obtains a classical 3D
Ising partition function in the limit of extreme anisotropy, namely 
couplings 
$2\beta
J_\parallel = -\log(\epsilon h)\gg 1$ along the imaginary time direction and
$2\beta J_\perp = \log((1+\epsilon)/(1-\epsilon))\ll 1$ along the spatial
directions, where the limit $\epsilon\to0$ needs to be taken. The two
order parameters can now both be understood as inserting a twist (i.e., coupling spins
antiferromagnetically) along a half-infinite plane. For our disorder
parameter, the boundary line of this plane is aligned along the imaginary
time axis, while for the one of Ref.~\cite{zhao:ising-disorder-operator}, it is alinged along a
spatial axis. Given the extreme anisotropy limit which needs to be
considered, it  is indeed plausible
that these two disorder operators would exhibit different scaling
behavior, as is  clearly confirmed by the numerics.

\subsection{What could $\beta^*$ be?}

A possible hypothesis for the value of $\beta^*$ in terms of the
underlying CFT could be based on assuming an effective description 
at the critical point where physical and virtual degrees of
freedom in the PEPS behave in the same way and thus exhibit the same
scaling of their correlations with an exponent $1+\eta$ (with $\eta$ the
anomalous dimension). Re-glueing the spins after twisting and integrating
over a finite cut suggests an exponent $\eta$ for a twist line correlator
at the critical point, and thus a critical exponent $\beta^*=\eta\nu/2$
for the order parameter away from criticality (where
$\beta=(1+\eta)\nu/2$); the resulting value 
$\beta^*\approx0.0114$ for the 3D Ising model agrees reasonably with the
magnitude of the observed value, given the difficulty to extract critical
exponents with high absolute precision. Indeed, this speculative formula
also matches the results obtained for topological phase transitions
observed in PEPS families which map to the (2+0)D Ising
model~\cite{iqbal:z4-phasetrans}, as well as the (2+0)D Ising model itself
(where $\beta=\beta^*=1/8$ due to the self-duality of the model) and mean
field (where $\beta^*=0$).

Could the critical exponent $\beta^*$ give us access to new universal
signatures of the phase transition?  First off, this depends on whether
$\beta^*$ can be derived from the underlying CFT at criticality, or more
generally the scaling exponents of the model at criticality. While this is
certainly plausible, the construction through twisting the PEPS ground
state -- which takes us outside the ground space! -- nevertheless leaves
the possibility that the critical exponent can, in fact, only be
obtained from the exponents of some extension of the
model.  In case $\beta^*$ can be derived from the scaling dimensions at
criticality, it will not give access to new information for the (2+1)D
Ising model, since the model if fully specified by two scaling dimensions
(which can be computed from $\beta$ and $\nu$). On the other hand, this
need no longer be true for more complex models with more scaling
dimensions, in which case case $\beta^*$ might give access to additional
universal data. Finally, even in case that the formula conjectured above
holds, or otherwise $\beta^*$ could be computed from $\beta$ and $\nu$, 
the exponent $\beta^*$ of the disorder operator still provides an
additional probe for universal behavior which, depending on the concrete
values of the exponents in a given scenario, might well allow to obtain
higher accuracy data about scaling dimensions as compared to other
exponents.

\subsection{Entanglement order parameters: A unifying perspective}

Let us conclude this section by explaining how topological order
parameters and the disorder parameters obtained from ``entanglement
twists'' can be understood on a unified footing as the most general order
parameters for tensor networks with symmetries. To this end, consider a
tensor with symmetry
\begin{equation}
\raisebox{-.65cm}{\includegraphics[scale=.9]{phys-sym-tensor}}
\end{equation}
where now, $U_g$ is \emph{not} necessarily a faithful representation --
therefore, this does, in particular, include the case of topological order
(for $U_g=\openone$), phases with physical symmetries exhibiting conventional
order ($U_g$ faithful), as well as symmetry enriched (SET) phases.

The most general order parameter for a tensor network with such a symmetry
should be an object which detects the breaking of any of those of the
symmetries, that is, an operator on either the physical or the entanglement
degrees of freedom which transforms as an irrep of the symmetry group.
However, placing such an irrep $S_\alpha$ on the \emph{physical} level can
always be replaced by placing a corresponding irrep $R_\alpha$ on the
virtual level, as 
\begin{equation}
\raisebox{-.5cm}{\includegraphics[scale=.9]{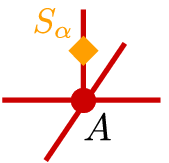}}
\mbox{\quad and\quad}
\raisebox{-.5cm}{\includegraphics[scale=.9]{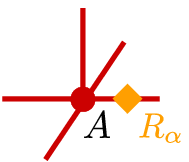}}
\end{equation}
transform in the same way. We thus find that the most general order
parameter is given by irreps acting on the ket and/or bra virtual indices
which transform as an irrep of the joint ket+bra symmetry group (depending
on the representation $U_g$, this can be $G$ (conventional order),
$G\times G$ (topological order), or something in between, as the physical
symmetry action has to cancel when building the ket-bra object). Similarly,
disorder operators can be constructed by strings of $V_g$ on the
entanglement or by membranes of $U_g$ on the physical degrees of freedom.
However, yet again, due to the relation shown in Fig.~\ref{fig:membrane}, any
physical symmetry membrane can be replaced by a string of $V_g$ on the
virtual layer. We thus find that the most general disorder parameter is
constructed from symmetry strings on the ket and/or bra layer of the
entanglement. As for the case of topological order parameters, order and
disorder parameters can be conbined, such as to detect symmetry protected
order (the most prominent example being string order parameters for 1D
symmetry protected phases).

We thus find that in all those cases, it is sufficient to construct
order/disorder parameters right away on the level of the entanglement
degrees of freedom, that is, as \emph{entanglement order parameters}.
From this perspective, the order parameters for topological and
for conventional order, including the new disorder parameter, are just
different manifestations of entanglement order parameters in settings with
different symmetry realization.

\section{Discussion\label{sec:discussion}}
\vspace*{-0.1cm}

Before concluding, let us discuss a few relevant aspects with regard to
our method.
\vspace*{-0.1cm}

\subsection{Gauge fixing}
\vspace*{-.1cm}

First, an interesting question is linked to the gauge fixing involved in
our algorithm. It can be easily checked that applying a random gauge
of the form \eqref{eq:peps-gauge-dof} independently for each point in the
phase diagram leads to a completely random and uncontrolled behavior of
the order parameter. Applying the gauge fixing procedure after such a
scrambling always returns the same tensor and thus stabilizes the behavior
of the order parameter again.  On the other hand, the data obtained in
numerical simulations typically does not display a random gauge; rather,
we expect the gauge to be determined by the choice of the initial tensor
and the way in which the optimization is performed (though this can of
course involve randomness or other effects which destabilize the gauge).
In particular, we have found that for data which has been obtained by
independently optimizing the tensor for the individual points in the phase
diagram, the optimized tensors yield an
order parameter with noticable residual noise, which can be significantly
improved by applying the gauge fixing procedure.  On the other hand, we
have also found that in order to obtain the best data, it is advisable to
initialize the tensors with the optimal tensors obtained for a nearby
point in the phase diagram (i.e., to adiabatically change the field); in
that case, we observe that the order parameters obtained from the 
optimized tensors already display a very smooth behavior, and applying an
additional gauge fixing step only leads to minor improvements.  This
is certainly plausible, given that an adiabatic change of the field only
leads to minor changes in the tensor and thus ideally to no significant
drift in the gauge.  

We have also compared different gauge fixing schemes (in particular,
the one described in Sec.~\ref{sec:subsec:recipe}, and a ``symmetric'' gauge fixing where
the spectrum of the left and right fixed point in Eq.~\eqref{eq:recipe:Ch-tilde} are fixed to be
equal), and found that they lead to slightly different order parameters,
which however display identical critical exponents, as expected.

\vspace*{-.1cm}
\subsection{Endpoints and vacua\label{sec:discussion-endpoints}}
\vspace*{-.1cm}

Second, the construction of our order parameters leaves open degrees of freedom
in the endpoint operators. On the one hand, in case of a trivial irrep
label, $\alpha=1$, there is no reason to restrict the endpoint to a single
irrep block $\gamma$ -- recall that we made this choice to obtain
gauge-invariant quantities when considering pairs of particle-antiparticle
endpoints -- since the gauge already cancels out for each endpoint
individually. We can thus replace $X_{\alpha,\gamma}$ in
Eqs.~\eqref{eq:recipe:N_g_a_g} and \eqref{eq:recipe:O_g_a_g} by any object
in the trivial irrep sector, that is, any $r = \sum w_\gamma
X_{\alpha=1,\gamma}$ (differently speaking, any $r$ with $V_grV_g^\dagger
= r$).  We have investigated this degree of freedom and found that while
it affects the (non-universal) value of the order parameter, it does not
affect the universal scaling behavior.

In addition, the endpoint operators
$X_{\alpha,\gamma}=\delta_{\gamma+\alpha,\gamma}\otimes M_{\alpha,\gamma}$
defined in Eq.~\eqref{eq:recipe:endpoint-ops} leave the freedom of
choosing different $M_{\alpha,\gamma}$ in the degeneracy space of the
irreps. Choosing different such $M_{\alpha,\gamma}$ can affect the
stability of the resulting curve (making a fitting of the scaling
difficult), where we have observed that our choice
$M_{\alpha,\gamma}=\openone$ leads to a particularly stable behavior. A
considerably more stable way of choosing $M_{\alpha,\gamma}$ different
from $\openone$ is
to impose that
$M_{\bar\alpha,-\gamma}=M_{\alpha,\gamma}^{-1}$, motivated by the fact that
this is the way how these two blocks transform relative to each other
under gauge transformations. Indeed, this yields more stable order
parameters (again with different values but the same scaling behavior),
but depending on the choice of $M$, we still observe cases where the curve
becomes unstable and does no longer allow for a reliable scaling analysis.
This suggests that the chosen gauge fixing is special, and changing the
gauge by a fixed invertible matrix can decrease the stability of the
method.

On the other hand, one might wonder whether one can also replace the
vacuum $X_{\mathrm{vac}}=\openone$ in
Eqs.~(\ref{eq:recipe:N_g_a_g}-\ref{eq:recipe:O_g_a_g}) by a different
operator describing an excitation in the trivial sector. We found that
this is not the case, as it can affect the observed critical behavior. This
might be understood as follows: An excitation in the trivial sector can be
seen as a particle-antiparticle pair; since each of those displays
critical behavior at the phase transition, we also expect -- and observe
-- such a non-analytical behavior for order-parameter-like quantities for
those trivial particles.  While these are not proper order parameters --
that is, they are non-zero on both sides of the transition -- they
nevertheless display a non-analyticity at the phase transition (similar to
the magnetization, cf.\ Fig.~\ref{fig:tc-V-VI-VII}). Thus, dividing the
order parameters by such a non-analytic normalization in
Eqs.~\eqref{eq:condfrac-recipe} and \eqref{eq:conffrac-recipe} will affect
the critical behavior in the regime where its non-analyticity 
dominates its non-zero value, and thus potentially mask the
true critical scaling.  We thus conclude that for the normalization, one
should use the trivial vacuum $X_\mathrm{vac}=\openone$.

\subsection{Why does it work at all?}

An interesting question one might raise is why the method works at all, and
why it gives meaningful results also in the trivial phase.

In particular, one might argue that if the PEPS optimization is carried
out with a very large bond dimension $D\to\infty$, one can easily
transform any iPEPS into one which additionally 
carries the entanglement symmetry \eqref{eq:G-injective}, yet 
without coupling this entanglement symmetry to the physics
of the system at all: To this end, simply take any PEPS with bond dimension $D$,
and construct a new PEPS with $D'=2D$ by tensoring each virtual index with
a qubit which is placed in the $\ket0$ state.  The new tensor has a
$\mathbb Z_2$ symmetry under the action of $\openone \otimes \sz$, while at
the same time, the additional virtual degree of freedom is completely
detached from the original PEPS, and thus can by no means give any
information whatsoever about the physics of the system.

The answer is that the finiteness of the bond dimension is relevant here
-- as long as the bond dimension is finite, using all degrees of freedom
is variationally favorable; in particular, it is favorable for the method
to use the symmetry-constrained degrees of freedom to encode the
topological degrees of freedom, as we have seen.  In that sense, going to
a large bond dimension -- where no energy is gained from the extra bond
dimension within numerical accuracy -- could in principle destabilize the
method, likely around a bond dimension $|G|D_\mathrm{crit}$, where
$D_\mathrm{crit}$ is the dimension where no further energy is gained in an
unconstrained optimization.
(E.g., for the 2D Ising model, it has been found that 
beyond $D_{\mathrm{crit}}=3$, variational optimization does not work
reliably any more due to the marginal gain in
energy~\cite{rader:peps-ising-scaling}; it is thus natural
to expect for the Toric Code $D_\mathrm{crit}=6$).

A related question is why the method still gives useful information in the
trivial phase, given that it probes the properties of topological
excitations.  This should, however, not come as a surprise: A phase
transition into an ordered phase (either conventionally, i.e.,
magnetically, or topologically ordered) is characterized by the formation
of ordered domains of increasing size $\xi$ which diverges at the phase
transition.  Thus, the structure of the ordered phase is already present
in the disordered phase sufficiently close to the transition, and thus,
using the entanglement symmetries to store this information is yet again
advantageous.  On the other hand, we have also seen in
Fig.~\ref{fig:tc-V-VI-VII} that for very large fields, where the
corresponding length scale becomes very small, and only a very small bond
dimension is needed for an accurate description of the ground state, the
data extracted from the entanglement degrees of freedom indeed starts to
become unstable and sensitive to initial conditions, with no effect on the
physical properties of the variational state, that is, it no longer
provides meaningful information about the system.

\subsection{Where does the additional order parameter come from?}

We have seen that in the Toric Code model, we were able to use our method
to construct an additional order parameter, which does not show up in the
(2+1)D Ising model.  This might come as a surprise, since there exists a
mapping from the ground state of the Toric Code model to that of the
(2+1)D Ising model. How can this be the case?

The explanation lies in the fact that by being constructed on the
entanglement degrees of freedom of the optimized ground state tensor, our
order parameters can leave the ground space of the Toric Code, and thus
the mapping to the ground space of the Ising model breaks down.  This has
been discussed in Sec.~\ref{sec:TCfield:mapping-to-ising}: Inserting
string operators at the entanglement level breaks loops, and the mapping
to the Ising model only works within the closed loop space.

As we have seen subsequently in Section~\ref{sec:opar-conventional}, this 
also opens up a new avenue for constructing order parameters based on PEPS
which is not restricted to topological order, by encoding physical
symmetries locally in the iPEPS tensor, and computing the response of the
wavefunction (i.e., the change in normalization) to the insertion of a
symmetry string along a cut at the entanglement level. Such a ``disorder
operator'' will show a distinct behavior in the two phases:  In the
ordered phase, where all degrees of freedom are aligned, it will give rise
to misaligned degrees of freedom all along the cut, and thus to a norm
zero.  On the other hand, in the disordered phase, the spins (and thus
tensors) are only correlated at the scale of the correlation length: The
misalignment along the cut will thus only persist for that distance, and
thus, a finite value of the order parameter is expected. 

In some sense, the ability of these order and disorder parameters to probe
otherwise unaccessible properties can be understood as emerging from the
interplay between the symmetry and entanglement structure of the
wavefunction with the local description enforced through the PEPS
description: This local description exposes the
way in which symmetries and entanglement build up locally, and thereby gives
access to information which cannot be simply extracted through local or
string-like operators acting on the physical degrees of freedom, as those
don't give access to information about how the quantum correlations in the
system organize locally.

In summary, PEPS with symmetries form a framework which allows to access
additional order parameters also for conventional phases, by optimizing
the iPEPS tensor and subsequently studying the response to symmetry twists
inserted on the entanglement level. They thus allow to extend disorder
parameters -- previously only defined for classical models at finite
temperature~\cite{kadanoff:disorder-operator,fradkin:disorder-operators}
-- to the domain of quantum phase transitions.

\section{Conclusions\label{sec:conclusion}}

In conclusion, we have presented a framework to construct and measure order
parameters for topologically ordered phases.  Our framework is based on
variational iPEPS simulations with a fixed entanglement symmetry, and the
ability of these symmetries to capture the behavior of anyons, and in
particular their disappearance at a phase transition through anyon
condensation and confinement. Importantly, we have devised methods to
construct and measure these order parameters in a gauge invariant way,
making the method suitable for fully variational iPEPS simulations where nothing
but the symmetry is imposed.  

We have applied our framework to the study of the Toric Code model in
simultaneous $x$ and $z$ fields, and have found critical exponents for
condensation $\beta$ and for the length scales associated with the mass
gap and confinement, $\nu$, which are consistent with the 3D Ising
universality class for the entire transition. In addition, however, our
method also allowed us to unveil a novel critical exponent for the order
parameter measuring the deconfinement fraction. 
 This
demonstrates the suitability of our framework for the microscopic study of
topological phase transitions.

We have then argued that our approach can, in fact, be understood more
generally as a way of defining order parameters using \emph{all}
symmetries present in the iPEPS tensors, leading to a general framework of
\emph{entanglement order parameters}, treating topological order and
global physical symmetries on a unified footing. In particular, we have
demonstrated that this allows to define novel disorder operators for
conventionally ordered phases such as the (2+1)D Ising model. We have
numerically studied the behavior of the disorder parameter for the latter
model at criticality, and found that it exhibits the same unknown critical
exponent as for the Toric Code above, demonstrating the power of the PEPS
framework and entanglement order parameters to probe critical behavior in
novel ways.

\begin{acknowledgements}
We acknowledge helpful comments by
E.~Fradkin,
S.~Gazit,
A.~Ludwig, 
F.~Pollmann.
S.~Rychkov, 
F.~Verstraete,
and W.-T.~Xu.
This work has received support from the European Union's Horizon 2020
program through the ERC-StG WASCOSYS (No.~636201) and the ERC-CoG SEQUAM
(No.~863476), and from the DFG (German Research Foundation) under Germany's
Excellence Strategy (EXC2111-390814868).
\end{acknowledgements}

\end{document}